%% file: mnras_tully_fisher_hengxing_mjj.tex
\documentclass[fleqn,usenatbib]{mnras}

\usepackage{times} 
\usepackage{listings}
\usepackage{graphics}
\usepackage{graphicx,xspace}
\usepackage{amsmath}
\usepackage{amssymb}
\usepackage{hyperref}
\usepackage{lscape}
\usepackage{rotating}
\usepackage{placeins}
\usepackage{natbib}
\usepackage{color}
\usepackage{txfonts}
\usepackage{mathptmx}
\usepackage{newtxtext,newtxmath}
\usepackage[export]{adjustbox}

\usepackage{bm}
\usepackage{caption}
\usepackage{subcaption}

\usepackage[dvipsnames]{xcolor}

\usepackage{grffile}

\newcommand{\ha}{H{\sc\,i}\xspace}

\newcommand{\vi}{\varv}
\newcommand{\simba}{{\sc Simba}}

\title[Measuring the baryonic Tully-Fisher relation below the detection threshold]{Measuring the baryonic Tully-Fisher relation below the detection threshold}

\author[H. Pan et al.]{Hengxing Pan$^{1,2,3}$\thanks{E-mail: hengxing.pan@physics.ox.ac.uk},
Matt J.~Jarvis$^{3,1}$,
Anastasia A.~Ponomareva$^{3}$,
Mario G. Santos$^{1,4}$,
James R. Allison$^{3}$,
\newauthor
Natasha Maddox$^{5}$ and Bradley S. Frank$^{4,6,7}$
\\
 $^{1}$Department of Physics and Astronomy, University of the Western Cape, Cape Town 7535, South Africa\\
 $^{2}$Purple Mountain Observatory, 2 West Beijing Road, Nanjing 210008, China\\
 $^{3}$Astrophysics, University of Oxford, Denys Wilkinson Building, Keble Road, Oxford OX1 3RH, UK\\
 $^{4}$South African Radio Astronomy Observatory (SARAO), 2 Fir Street, Observatory, 7925, South Africa\\
 $^{5}$Faculty of Physics, Ludwig-Maximilians-Universit{\"a}t, Scheinerstr. 1, 81679 Munich, Germany\\
 $^{6}$The Inter-University Institute for Data Intensive Astronomy (IDIA), and
University of Cape Town, Private Bag X3, Rondebosch, 7701, South Africa\\
 $^{7}$Department of Astronomy, University of Cape Town, Private Bag X3, Rondebosch 7701, South Africa
}
\date{Accepted XXX. Received YYY; in original form ZZZ}

\pubyear{2019}

\begin{document}
\label{firstpage}
\pagerange{\pageref{firstpage}--\pageref{lastpage}}
\maketitle

\begin{abstract}
We present a novel 2D flux density model for observed \ha emission lines combined with a Bayesian stacking technique to measure the baryonic Tully-Fisher relation below the nominal detection threshold. We simulate a galaxy catalogue, which includes \ha lines described either with Gaussian or busy function profiles, and \ha data cubes with a range of noise and survey areas similar to the MeerKAT International Giga-Hertz Tiered Extragalactic Exploration (MIGHTEE) survey. With prior knowledge of redshifts, stellar masses and inclinations of spiral galaxies, we find that our model can reconstruct the input baryonic Tully-Fisher parameters (slope and zero point) most accurately in a relatively broad redshift range from the local Universe to $z = 0.3$ for all the considered levels of noise and survey areas, and up to $z = 0.55$ for a nominal noise of $90$\,$\mu$Jy/channel over 5 deg$^{2}$. Our model can also determine the $M_{\rm HI} - M_{\star}$ relation for spiral galaxies beyond the local Universe, and account for the detailed shape of the \ha emission line, which is crucial for understanding the dynamics of spiral galaxies. Thus, we have developed a Bayesian stacking technique for measuring the baryonic Tully-Fisher relation for galaxies at low stellar and/or \ha masses and/or those at high redshift, where the direct detection of \ha requires prohibitive exposure times. 

\end{abstract}




\section{Introduction}

The Tully-Fisher relation (TFr, \citealt{tf77}) is  an empirical relationship that links the luminosity of a disk galaxy to its rotational velocity \citep{sorce14,ponomareva18, Lelli19}. 
The TFr remains one of the most studied dynamical scaling relations of disk galaxies, and has been shown to hold over a wide wavelength range \citep{ponomareva17}, in different environments \citep{willick99, Abril-Melgarejo2021} and for galaxies of different morphological types   \citep{Chung2002, courteau03, Bedregal2006, Karachentsev2016}. The TFr was first established as a powerful tool to make redshift-independent distance estimates for spiral galaxies, with the goal of deriving the local peculiar velocity field \citep{tully14}.

A more elementary form of the TFr is the so-called baryonic Tully-Fisher relation (bTFr), which relates the baryonic component of a galaxy (characterised by its total baryonic mass) to its total dynamical mass (including the DarkMatter; DM), characterised by the rotational velocity. Although the exact nature of the bTFr remains unclear, it is one of the most fundamental relations in the framework of galaxy formation and evolution since it places strong constraints on the properties of galaxies produced by cosmological simulations (e.g. \citealt{glowacki20, Dubois2020}). 

One of the open questions in the comparison of these simulations and observations is the presence or absence of curvature at the low-mass end of the bTFr. While curvature is a generic prediction of the models \citep{Trujillo-Gomez2010,Desmond2012}, because it arises from the need to achieve low baryon fractions for both high- and low-mass haloes \citep{Papastergis2012}, it has not been observed even in the bTFr studies of heavily gas dominated galaxies \citep{Begum2008,Pap2016, iorio2017}. However, since the low-mass ($\lesssim10^9$ M$_\odot$) end of the bTFr is always poorly constrained due to the faintness of the atomic hydrogen (\ha) signal which usually lies below the detection threshold (\ha being the main mass contributor), there is still no definitive answer regarding the slope of the bTFr at this mass range.

At low redshift ($z<0.1$), \ha is used as a kinematic tracer for the bTFr studies as it extends far beyond the optical radius, better tracing the whole of the DM halo. At higher redshifts carbon monoxide (CO) and optical lines (e.g. H$\alpha$, [O{\sc iii}]) have been used as kinematic tracers for studying the evolution of the bTFr \citep{ubler17, topal18,tiley19}, since the \ha emission line has for a long time remained undetectable beyond the local Universe due to its intrinsic faintness \citep{fernandez2016,Chowdhury_2020}. The study of the bTFr evolution is extremely important since it can provide valuable insights on the mechanisms of galaxies' growth and  help place further constraints on the cosmological simulations of galaxy formation and evolution. Significant evolution of the bTFr would point to the imbalance in the mass assembly of the baryons and the DM haloes due to the effect of the halo on the formation of the central galaxy. 
For example, a recent study of the state-of-the-art cosmological simulations \simba\ \citep{Dave2019} found a clear change in the visual distribution, as well as the linear best-fit parameters to the bTFR over the redshift range between $\rm z=0$ and $z=1$, mostly due to the differences in the merger histories of the DM haloes \citep{Glowacki2020}. In that work, they make predictions for the future \ha surveys which need to be tested.

To date, there has been no definitive observational conclusion regarding the evolution of the bTFr, due to the different kinematic tracers and samples used. Ideally one would need to compare the statistical properties of the \ha-based bTFr (slope, scatter and zero point) at different redshifts, as well as to use similar  methods to measure these properties \citep{bradford16}. Unfortunately, this remains intractable even with the modern state-of-the-art observational facilities. For example, the \ha data from the MIGHTEE survey \citep{jarvis2017meerkat, Maddox2020}, which is being undertaken with the Meer Karoo Array Telescope (MeerKAT, \citealp{Jonas2016}), will not detect individual galaxies with a typical \ha gas mass ($\sim M_{\rm HI}^{*}$)
beyond $z\sim 0.3$ \citep{Maddox2020}. Thus, to understand the evolution of the bTFr, and hence the evolution of gas in galaxies and its gravitational interplay with the DM haloes, one way to progress forward is to use stacking techniques in order to extend the accessible redshift and mass range by measuring a statistical signal \citep{rhee2018, delhaize2013detection}. 

\citet{Meyer2016} showed that the TFr can be recovered via the spectral stacking technique by comparing the statistical properties of the stacked K-band TFr with the direct measurements from HIPASS \citep{zwaan2005hipass}. Moreover, they showed that no knowledge of the individual galaxy inclinations, or even the inclination distribution of the data is needed to accurately recover the properties. However, this technique is based on the classic stacking approach, where the spectral line data at the known locations and redshifts of many galaxies are co-added to improve the signal-to-noise ratio, thus losing information on individual galaxies contributing to the stacked spectra by averaging out their properties.

In this paper we develop a  novel approach and extend the Bayesian stacking technique, which was introduced for measuring  the \ha Mass Function (HIMF) below the detection threshold \cite[][hereafter P20]{pan2020}, to the measurement of the bTFr. This work provides the framework to measure the \ha-based baryonic Tully-Fisher relation beyond the local Universe, including low \ha mass galaxies. 

This paper is organised as follows. In Section \ref{sec:method}, we describe our simulated galaxy catalogues and \ha data cubes. In Section \ref{sec:model}, we detail our method for modelling the 2D flux density and present an extended Bayesian stacking technique based on P20. We show the results of implementing this technique to our simulated data sets in Section \ref{sec:results} and present out conclusions in Section \ref{sec:conclusions}. Throughout the paper, we adopt a concordance cosmology with a Hubble constant $H_{0}=67.7$ km$\cdot$s$^{-1}\cdot {\rm Mpc}^{-1}$, total matter density $\Omega_{\rm m}=0.308$ and dark energy density $\Omega_{\Lambda}=0.692$ \citep{ade2016planck}.

\section{Simulated Data}
\label{sec:method}

In order to test our method to measure the bTFr below the nominal detection threshold of \ha galaxies in a survey, we first require a simulated \ha data cube populated with galaxies of known baryonic mass. To simulate such a cube, we need to construct the \ha 21-cm line flux distribution per frequency channel assuming reasonable emission-line profiles, a source-count model and the expected noise properties for a typical survey.

\subsection{Generating the galaxy sample}
\label{sec:himf}

First, we simulate an optical catalogue of galaxies with known stellar masses using an assumed galaxy Stellar Mass Function (SMF). The SMF represents the intrinsic number density of galaxies in the Universe as a function of their stellar mass at a given redshift. We use a Schechter function model \citep{Schechter1976} along with a pure density evolution term to characterise the evolution of the SMF with redshift:
\begin{equation}
    \phi(M_{\star}, z) = \ln(10) \space \phi_\ast\left(\frac{M_{\star}}{M_\ast}\right)^{\alpha_\ast + 1} e^{-\frac{M_{\star}}{M_\ast}} (1+z),
    \label{eq:himf}
\end{equation}
where $\phi_\ast = 1.5\times10^{-3}$, $M_\ast=1.3\times10^{11}$\,M$_\odot$ and $\alpha_\ast=-1.1$ correspond to the normalisation, characteristic mass and faint-end slope, respectively.

We then use the assumed SMF to generate a sample of simulated galaxies over a volume similar to the one expected for the MIGHTEE-\ha survey (i.e. $\approx 20$\,deg$^2$ over the redshift range $0<z<0.55$). With the big survey volume and exceptional multi-wavelength ancillary data, MIGHTEE-\ha is an ideal testbed for our novel approach for constraining the bTFr. We note that the exact form of the SMF does not impact on our results, although in reality an accurate account of the stellar-mass incompleteness for the parent galaxy sample may be needed to account for any possible biases that arise from sample selection.

\subsection{HI and stellar mass relation of spiral galaxies}

\begin{figure}
  \centering
    \includegraphics[width=\columnwidth]{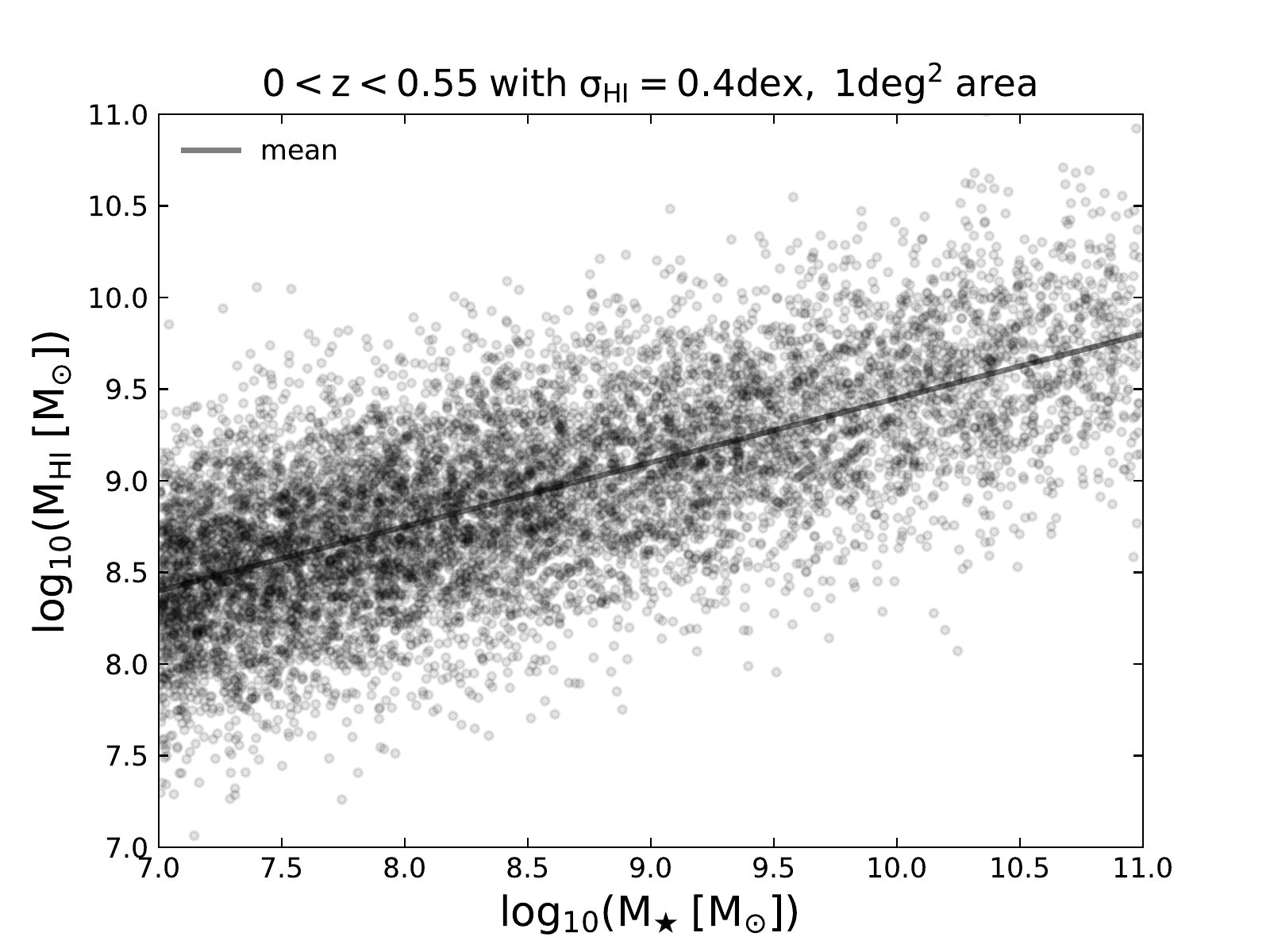}
 \caption{Simulated HI and stellar mass relation with $\sigma_{\rm HI}$ =0.4 dex on the \ha mass.}\label{fig:hi-star}
\end{figure}

The relationship between the \ha and stellar mass in galaxies is not straightforward. For example, it was shown by \citet{Maddox2015} that it is not linear and tends to flatten out at higher stellar masses. \citet{Parkash2018} considered various morphological samples of galaxies, and found that for a sample of spiral galaxies this relationship can be described as
\begin{equation}
    \log_{10}(M_{\rm HI}) = 0.35[\log_{10}(M_{\star})-10]+9.45,
    \label{eq:HI-star}
\end{equation}
with 68\% of the \ha masses within 0.4 dex uncertainty ($\sigma_{\rm HI}$), where $M_{\rm HI}$ and $M_{\star}$ are the galaxy \ha and stellar mass, respectively. 
Thus, we generate the population of spiral galaxies with \ha masses defined as
\begin{equation}
    \log_{10}(M_{\rm HI}) = 0.35[\log_{10}(M_{\star})-10]+9.45 + \mu,
    \label{eq:HI-star_scatter}
\end{equation}
where $\mu$ is a random variable that follows a normal distribution with mean 0 and standard deviation $\sigma_{\rm HI}$=0.4 dex.

\begin{figure}
    \includegraphics[width=\columnwidth]{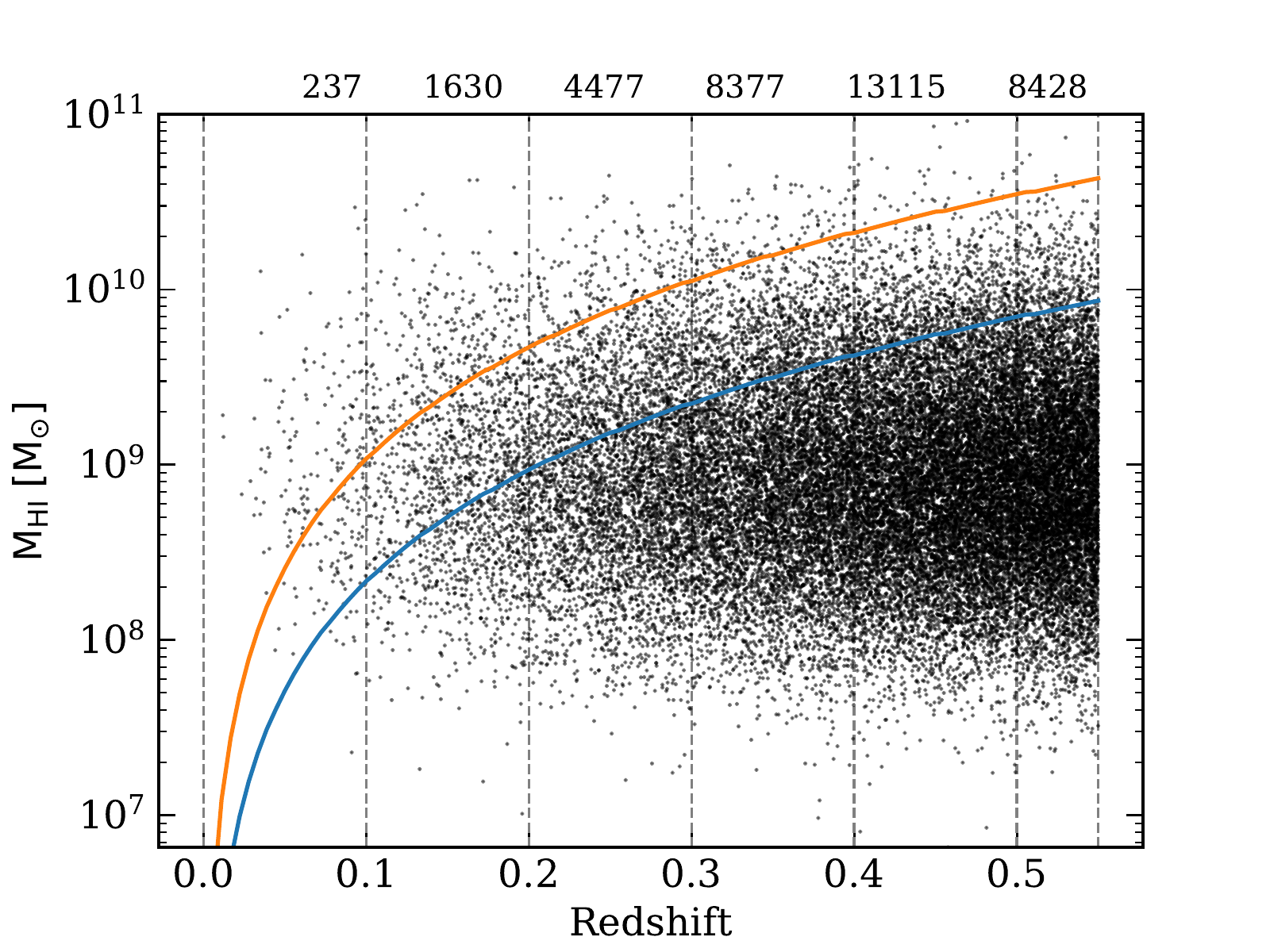}
    \caption{Simulated \ha mass versus redshift over a 1\,deg$^2$ area with a scatter of $\sigma_{\rm HI}$ =0.4 dex on the \ha mass, and lower and upper stellar mass limits of $M_\star = 10^7$ and $10^{11}$\,M$_\odot$. The colour-coded lines indicate the $\sigma_{\rm n} = \sqrt{N_{\rm ch}}\sigma_{\rm ch}\vi$ and 5$\sigma_{\rm n}$ detection threshold. The numbers listed at the top of the figure indicate the number of galaxies in each $\Delta z=0.1$ redshift bin.}
    \label{fig:mass_z}
\end{figure}

Figure~\ref{fig:hi-star} shows the simulated \ha mass as a function of stellar mass, with lower and upper stellar mass limits of $M_\star = 10^7$ and $10^{11}$\,M$_\odot$ for our simulated survey. The stellar mass range we use is applicable to most \ha-bearing galaxies. Of course, a colour or morphological selection would help to better determine the \ha masses, however this selection mostly affect the highest mass objects, which do not change our results significantly due to their low numbers as we will demonstrate in Section~\ref{sec:results}. Figure~\ref{fig:mass_z} shows the \ha mass as a function of redshift, from which it is clear that the vast majority of \ha galaxies lie below the nominal noise threshold of the MIGHTEE survey (indicated by the coloured lines). Our goal here is to take these faint \ha galaxies into account.

We note that the assumed HI and stellar mass relation is rather simplistic than realistic as more low-HI-mass galaxies are expected from HI surveys and the scatter on this relation may be mass dependent. A more sophisticated model, such as a 2D distribution of galaxies as a function of HI and stellar mass (or even extended to a 3D distribution if the velocity width is also taken as a variable), could be implemented to better quantify the $M_{\rm HI}-M_{\star}$ relation. However, this would just add a few more free parameters to our model and will not influence the main results as the majority of spiral galaxies follows Eq.~\eqref{eq:HI-star} well. At least, over a range of masses $9<\log_{10}(M_{\rm HI}/M_\odot)<11$, the $M_{\rm HI}-M_{\star}$ relation is fairly linear, and at masses lower than that, the constraining power on the bTFr is less strong as we will see in Figure~\ref{eq:tully-fisher}. Although a more complex treatment could be applied in future to real data, this is beyond the scope of this paper where we wish to highlight the key features of the method.

\subsection{The baryonic Tully-Fisher relation}
To estimate the total baryonic mass of our simulated galaxies we calculate the total mass of a galaxy disk as $M_{\rm disk} = M_{\star} + M_{\rm gas}$, assuming $M_{\rm gas} = 1.4M_{\rm {\sc HI}}$ to account for the presence of helium and metals \citep{A1996book}. We can then predict the rotational velocity of a galaxy ($\rm V_{rot}$) using an assumed baryonic Tully-Fisher relation:
\begin{equation}
    A \left(\frac{W}{200~{\rm km/s}}\right)^\alpha = M_{\rm disk},
    \label{eq:tully-fisher}
\end{equation}
where $\rm W=2V_{rot}$ is the corrected width of the \ha line profile, and we take the normalisation (i.e. zero point) $A=2 \times 10^9$ and the slope $\alpha = 4$ as our canonical model \citep{McGaugh2000}.
We simulate the inclination $\sin(i)$ as uniformly distributed from 0 to 1, so the observed velocity width ($W_{50}$) is defined as:
\begin{equation}
    W_{50} = W\times \sin(i).
    \label{eq:inclination}
\end{equation}

\subsection{Profiles of the \ha emission lines}
To determine whether we can measure the bTFr below the noise threshold, we need to simulate realistic \ha line profiles. Therefore, we adopt a Gaussian profile to describe the shape of the \ha flux density distribution of dwarf and face-on galaxies:
\begin{equation}
    G(v) = \frac{a}{\sigma\sqrt{2\pi}}\times e^{-\frac{(v-v_0)^2}{2\sigma^2}},
    \label{eq:gau}
\end{equation}
where $\sigma = W_{50}/2.3548$, $v_0$ is the central frequency at a given redshift, and $a$ is the integrated flux determined by the \ha mass and redshift. 
Whereas for inclined spirals, we use the busy function \citep[BF,][]{Westmeier2014} for incorporating the double-horn profiles:
\begin{multline}
    B_1(v) = \frac{a}{4}
            \times (\mathrm{erf}[b_{1} \{ w + v - x_{\rm e} \} ] + 1) \\
            \times (\mathrm{erf}[b_{2} \{ w - v + x_{\rm e} \} ] + 1)
            \times ( c|v - x_{\rm p}|^{n} + 1),
    \label{eq:b1}
\end{multline}
where $w$ is the half-width of the observed profile; $w \sim W_{50}/2$. We also find that setting the parameter $c = (r_{\rm c}-1) w^{-n}$ allows for more realistic profiles to be generated for broad lines. This relation defines a new fixed parameter $r_{\rm c}$, replacing the original parameter $c$ and describing the ratio of peak flux to the trough flux, as we find the peak flux of a double-horn line is approximately proportional to the flux at $|v - x_{\rm p}| = w$ for broad lines. We note that $r_{\rm c}$ is only an indicator of that ratio rather than a precise definition. Nevertheless, it provides an intuitive way of controlling the shape of generated lines by the busy function. The probability distribution of these input parameters for generating the \ha lines with BF profiles are listed in Table~\ref{tab:busy}.

We limit the BF profiles to the more edge-on galaxies with the velocity widths larger than 120~km/s. Given the velocity width per channel in our cubes $\sim 5.5$~km/s at $z = 0$, and at least a few channels are needed to resolve the double horns on the line profile, we chose the following specific criteria for simulating our \ha lines:
\begin{equation}
    S(v) =
    \begin{cases}
        B_1(v)& \text{for}~\sin(i) \ge 0.5 \text{ and } W_{50} \ge 132 {\rm km/s}, \\
        \;\:G(v)  & \text{otherwise}. \\
    \end{cases}
    \label{eqn_obreschkow}
\end{equation}

\begin{table}
\centering
\caption{The parameters of the busy function for simulating the \ha lines. We note that only parameters $b_{1}, b_{2}, r_{\rm c}$ and $x_{\rm p}$ are randomly sampled. The centroid of the error function, $x_{\rm e}$, is fixed at the redshift of the source. The parameter $r_{\rm c}$ is the original parameter $c$ weighted by the half-width $w$.}
\label{tab:busy}
\begin{tabular}{lll}
\hline
Parameter &  Meaning & Probability Distribution \\
\hline
$b_{1}, b_{2}$             &steepness of line flanks & uniform $\in [0.3, 0.7] $ \\
$r_{\rm c}$                      &ratio of peak to trough& uniform $\in [1.5,    2]$ \\
$|x_{\rm p}-x_{\rm e}|$    &difference of centroids & uniform $\in [0, 0.3w] $\\
\hline
\end{tabular}
\end{table}

We also need to determine the velocity width $W$ given by the baryonic Tully-Fisher relation (Eq. \ref{eq:tully-fisher}) and the amplitude $a$ fixed by the integrated flux $S =\int S(v) dv$. 
The integrated flux can be converted from the \ha mass under the usual assumptions (i.e. optically thin gas, \citealt{meyer2017tracing}) via
\begin{equation}
    M_{\rm HI} = 2.356 \times 10^5 D^2_L(1+z)^{-1} S,
	\label{eq:factor}
\end{equation}
where the \ha mass ($M_{\rm HI}$) is in units of solar masses, the luminosity distance to the galaxy ($D_L$) is in Mpc, and the integrated flux ($S$) is in Jy\,km\,s$^{-1}$. The ($1 + z$) factor is needed when $S$ is expressed in units of Jy\,km\,s$^{-1}$ rather than Jy\,Hz.

Finally, we populate the simulated cubes with the emission lines, including Gaussian noise with standard deviation similar to that expected for the MIGHTEE survey. Specifically, we adopt $\sigma_{\rm ch} = 90$\,$\mu$\,Jy/channel, which is the expected noise per $26$\,kHz channel for MIGHTEE. The simulated cubes have a spatial resolution of $5$\,arcsec and a total of 19504 spectral channels from 913.4\,MHz to 1420.5\,MHz. We note that the real cubes from the new generation of \ha surveys are likely to have lower spatial resolution than this, and source confusion may increase the noise slightly, however here we demonstrate the overall methodology.

\begin{figure}
 \includegraphics[width=1.\columnwidth]{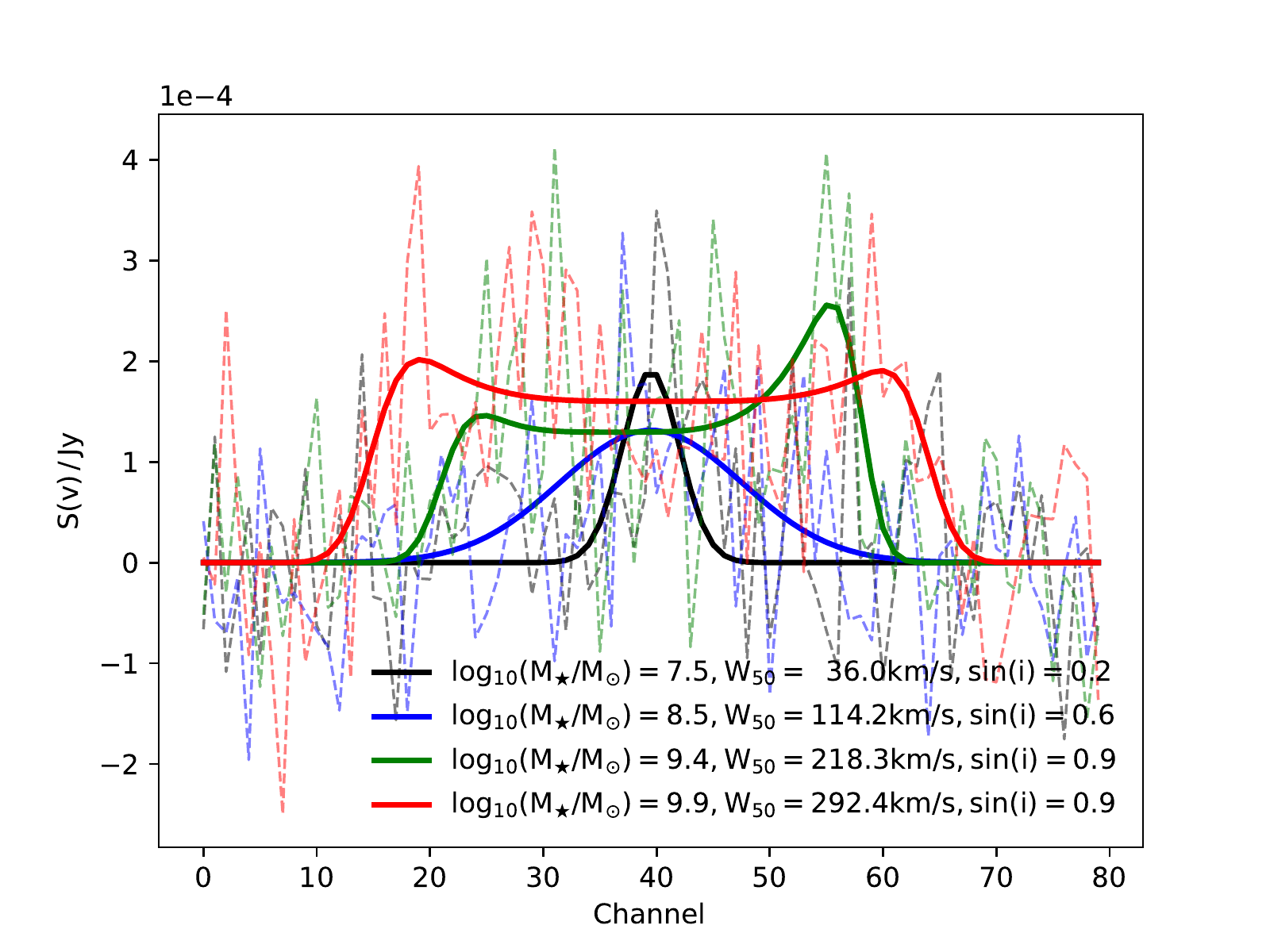}
 \caption{Examples of flux density as a function of channel at $0.1 < z < 0.2$. The solid lines are pure signals while the dashed lines are signals added to the noise. The central channels of the spectra are fixed at channel 40.}\label{fig:spectra}
\end{figure}

The measured flux from the \ha emission line $S_m(v)$, is a combination of the intrinsic flux from the source $S(v)$ plus the contribution from the noise $S_n(v)$:
\begin{equation}
    S_m(v) = S(v)  + S_n(v),
    \label{eq:gau}
\end{equation}
where $S_n(v) = \rm Normal(0,\; \sigma_{\rm ch})$ for this simulation. Figure~\ref{fig:spectra} shows a few examples of flux density that have been redshifted to $0.1 < z < 0.2$ as a function of the frequency channel with mixed Gaussian and BF profiles. As expected from the bTFr, and from the assumed $M_{\rm HI}-M_{\star}$ relation, the $M_{\star}$ mass correlates with the line width when corrected for the galaxy inclination. 

\section{The Statistical Model}
\label{sec:model}

Being able to simulate the emission lines in essence indicates that we can build a model to determine the bTFr for \ha galaxy surveys. Even though we will not be able to measure the \ha mass of individual galaxies below the nominal detection threshold, for surveys like MIGHTEE, which have excellent ancillary data, we will have information on the stellar masses, inclinations and redshifts. With the shape of the $M_{\star}-M_{\rm HI}$ relation, knowing the $M_{\star}$ of a galaxy gives us a predicted probability density function (PDF) for the $M_{\rm HI}$ (i.e. to generate \ha samples shown in Figure~\ref{fig:hi-star}), which leads to the PDFs for the integrated flux ($S$), gas mass ($M_{\rm gas} = 1.4M_{\rm HI}$) and therefore the disk mass ($M_{\rm disk} = M_{\star} + M_{\rm gas}$). 

Based on the assumed bTFr, we can further predict the PDF for the velocity width of the \ha line profile of a single galaxy, and for the integrated flux that determines the amplitude. Along with the assumed \ha line profiles, we can then predict the PDF for the \ha flux density of the galaxy in each frequency channel. Conversely, if we do not know the parameters $A$ and $\alpha$ for the bTFr, we can then constrain them with the observed \ha emission lines in return, which is the aim of this work.

\subsection{Modelling the 2D flux density}

Given our model, the probability of having the measured fluxes for a single source,  $P(S_{\rm m}(v))$, can be expressed as
\begin{equation}
P(S_{\rm m}(v)) = \int d M_{\rm HI}P(M_{\rm HI}) \prod_{v} P_n(S_{\rm m}(v)-S(v, M_{\rm HI})),
\label{eq:probability}
\end{equation}
where $P(M_{\rm HI})$ is the PDF for the $M_{\rm HI}$ at a fixed stellar mass, and $P_n$ follows the noise distribution of $\rm Normal(0,\; \sigma_{\rm ch})$ in each channel independently from other channels. $S(v, M_{\rm HI})$ is the modelled 2D flux density as a function of channel number and \ha mass. However, the relationship between the \ha mass and flux density $S(v)$ is non-linear, thus the $S(v, M_{\rm HI})$ must be solved numerically.

In fact, we model the $S(v, M_{\rm HI})$ for each channel by generating a distribution of \ha samples for a given stellar mass, and then propagating them into the final flux density as we generate the signals. We use the quantities with a tilde on the top of their symbols as the samples that meet a given distribution, e.g. $\log_{10}(\widetilde{M}_{\rm HI})$ indicates a group of \ha samples with their mass following a distribution of $\rm Normal\left(\log_{10}(M_{\rm HI}), \; \sigma_{\rm HI}\right)$.

If we only take into account the uncertainties of emission lines caused by the scatter on the $M_{\rm HI} - M_{\star}$ mass relation, then the model for predicting the velocity width $W_{50}$ and the integrated flux $S$ is composed of the following equations: 
\begin{equation}
    \log_{10}(\widetilde{M}_{\rm HI}) = 0.35[\log_{10}(M_{\star})-10]+9.45,
    \label{eq:HI-distribution}
\end{equation}

\begin{equation}
    \widetilde{M}_{\rm disk} = M_{\star} + 1.4\widetilde{M}_{\rm HI},
    \label{eq:disk_mass_model}
\end{equation}

\begin{equation}
    A \left(\frac{\widetilde{W}}{200 {\rm km/s}}\right)^\alpha = \widetilde{M}_{\rm disk},
    \label{eq:tully-fisher_model}
\end{equation}

\begin{equation}
    \widetilde{W}_{50} = \widetilde{W}\times \sin(i),
    \label{eq:inclination_model}
\end{equation}

\begin{equation}
    \widetilde{S} = \widetilde{M}_{\rm HI}/(2.356 \times 10^5 D^2_L(1+z)^{-1}).
	\label{eq:factor_model}
\end{equation}

For modelling the Gaussian lines, we take $S(v) = G(v)$. For  the double-horn lines, we adopt a simplified version of the BF with six free parameters as
\begin{multline}
    B_2(v) = \frac{a}{4}
            \times (\mathrm{erf}[b\{ w + v - x\} ] + 1) \\
            \times (\mathrm{erf}[b\{ w - v + x\} ] + 1)
            \times ( c|v - x|^{n} + 1).
    \label{eq:b2}
\end{multline}
Here we replace $b_1, \;b_2$ with $b$, and $x_{\rm p}, \;x_{\rm e}$ with $x$ compared with Eq.~\eqref{eq:b1}. If we set  $n=4$ and $c = (r_{\rm c}-1)w^{-n}$, along with the predicted integrated flux and velocity width, the final free parameters of $B_2(v)$ are reduced to be the steepness of the line flanks $b$ and ratio of peak to trough $r_{\rm c}$. We then take these two parameters as nuisance variables in the model.

\begin{figure}
  \centering
    \includegraphics[width=1.1\columnwidth]{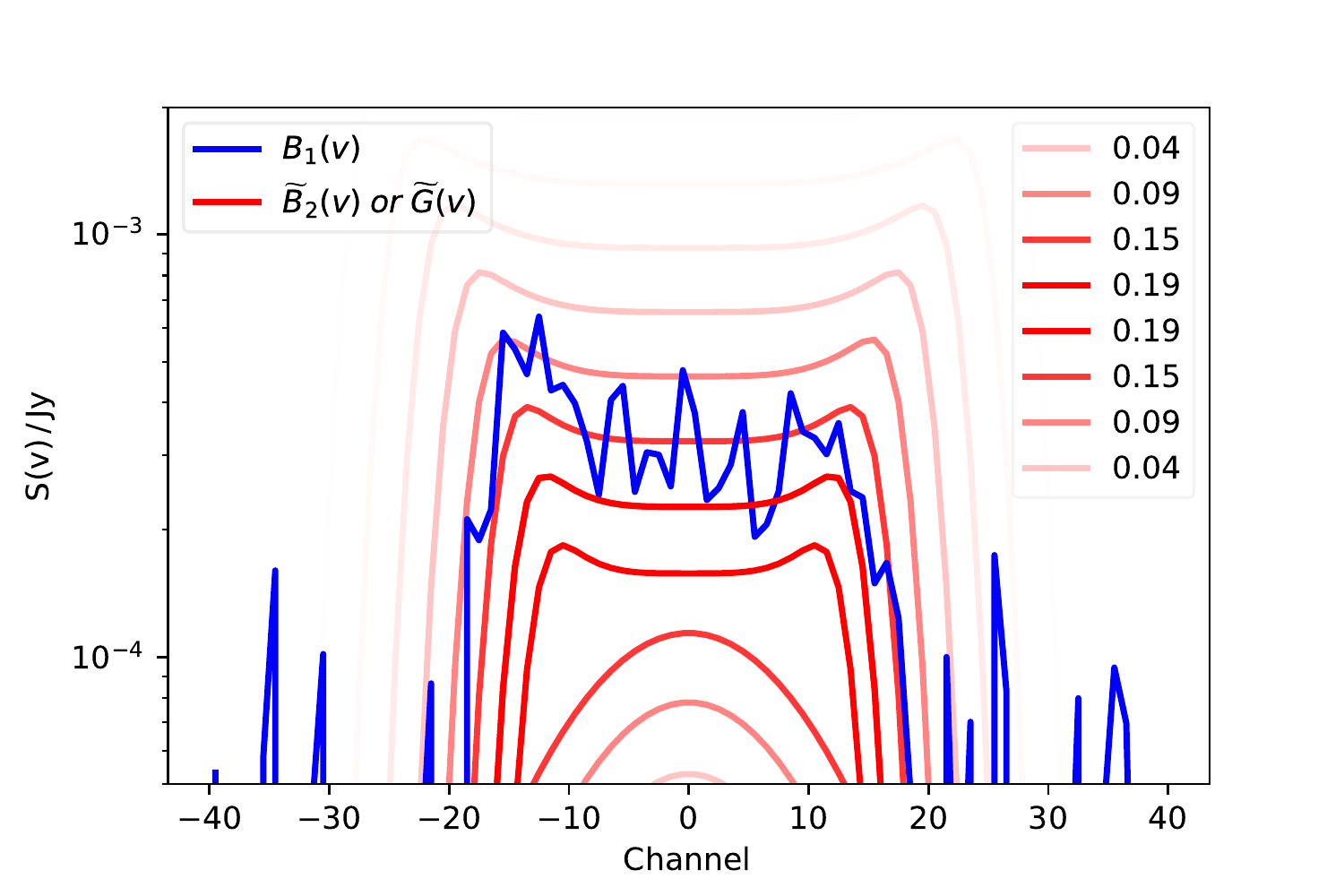}
 \caption{An example of modelling the asymmetric $B_1$ profile with a 2D flux density model as a function of channel number and \ha mass. The blue line is the generated \ha emission line added to the noise at $z\sim0.06$. The red lines are the modelled flux densities with a range of \ha masses corresponding to their probabilities indicated in the right legend, including line profiles of a symmetric busy function $B_2$ and Gaussian function.}\label{fig:B1B2}
\end{figure}

From Eq.~\eqref{eq:b2}, it can be found that $B_2(v)$ is symmetric once we have $b_1=b_2$ and $x_{\rm p}=x_{\rm e}$. The aim is to make a simplified form  ($B_2$) of the busy function that can model the width and amplitude of the busy function $B_1$ for the simulated lines. Based on this equation, we can then model the intrinsic flux densities on each frequency channel for the simulated \ha galaxy samples as
\begin{equation}
    \widetilde{S}(v) =
    \begin{cases}
        \widetilde{B}_2(v)& \text{for}~\sin(i) \ge 0.5 \text{ and } \widetilde{W}_{50} \ge 132 {\rm km/s} , \\
        \;\:\widetilde{G}(v)  & \text{otherwise}. \\
    \end{cases}
    \label{eqn_obreschkow}
\end{equation}

In Figure~\ref{fig:B1B2}, we show an example of the $B_1(v)$ added to the channel noise in blue for a galaxy with  $\log_{10}(M_{\star}/M_\odot) =8.15$ and $\log_{10}(M_{\rm HI}/M_\odot)=9.14$, and the modelled $\widetilde{B}_2(v)$ and $\widetilde{G}(v)$ profiles in red with a range of \ha mass centred at the averaged \ha mass of $\log_{10}(M_{\rm HI}/M_\odot)=8.8$ from Eq.~\eqref{eq:HI-distribution}. The transparency of red lines on the right legend indicates the probability of having the corresponding \ha mass. Since the averaged \ha mass in our model is slightly lower than the simulated \ha mass for the survey, the $B_2(v)$ profile with a maximum probability of 19\% is also slightly below the observed (blue) line. With higher or lower \ha masses than the $\log_{10}(M_{\rm HI}/M_\odot)=8.8$, the probabilities get smaller as expected. In particular, the double-horn profile $B_2(v)$ will change to a Gaussian profile $G(v)$ when the line width $W_{50}$ is less than 132~km/s. We emphasise that the parameters $b$ and $r_{\rm c}$ for the $B_2$ profile are set to be the mean value of the input ranges, while parameters $b_1, b_2$ and $r_{\rm c}$ for the $B_1$ profile are randomly assigned. 
\\

We note that the $\widetilde{S}(v)$ is essentially the 2D flux density $S(v, M_{\rm HI})$ that we are trying to solve in Eq.~\eqref{eq:probability}, and there are in total five parameters for modelling the flux densities for a given survey, three ($A, \alpha$ and $\sigma_{\rm HI}$) of which are used for characterising the assumed bTFr and $M_{\rm HI} - M_{\star}$ relation, with the extra two ($b$ and $r_{\rm c}$) to further describe the shape of the measured \ha lines. One may wonder why we need these two nuisance variables to model these lines, since we can only predict the line strengths and widths. The point is that modelling the dip of the \ha line between the double horns is an important element for our model to be accurate. None of the simpler functions, like Gaussian or Top-hat profiles, are able to characterise this dip, therefore, they will inevitably cause a model bias.

In conclusion to this section, we emphasise that the prior knowledge of stellar masses, inclinations and redshifts of the HI galaxy samples is required for modelling these 2D flux densities effectively. In reality this data may not be accurate or complete, and eventually more uncertainties may need to be introduced. That is precisely why we are now developing this approach and intend to implement it in the MIGHTEE fields, where exceptional multi-wavelength ancillary data are available. Of course, there may also be non-Gaussian noise or correlated noise in the real HI data, and we can address this issue with a fitted noise distribution to the real data as mentioned in P20. Nevertheless, this method is a useful first step towards the application to real data.

\begin{figure*}
  \centering
  \begin{subfigure}[b]{0.33\textwidth}
    \includegraphics[width=1.1\columnwidth, height=0.86\columnwidth]{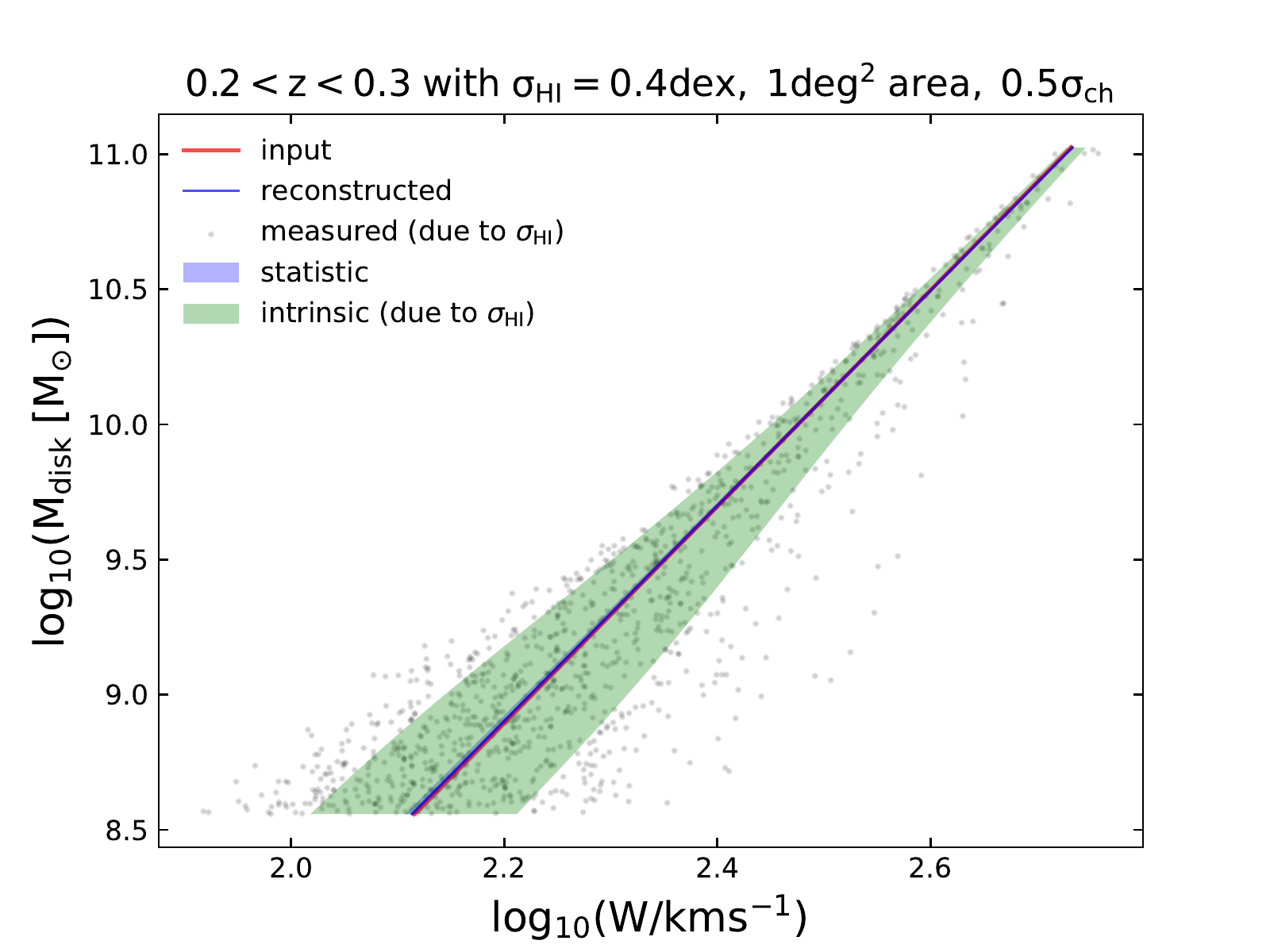}
    \includegraphics[width=1.1\columnwidth, height=0.86\columnwidth]{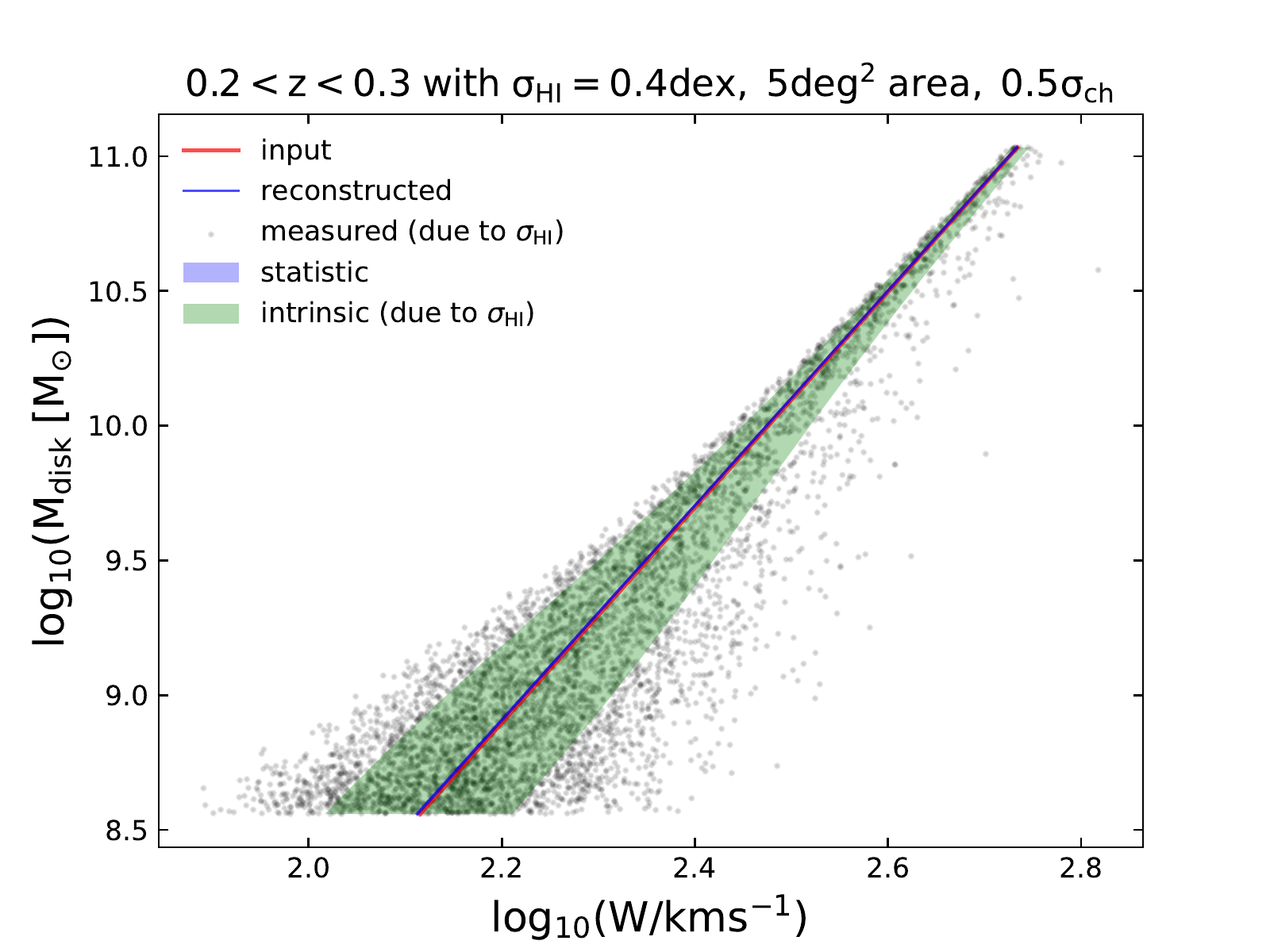}
    \includegraphics[width=1.1\columnwidth, height=0.86\columnwidth]{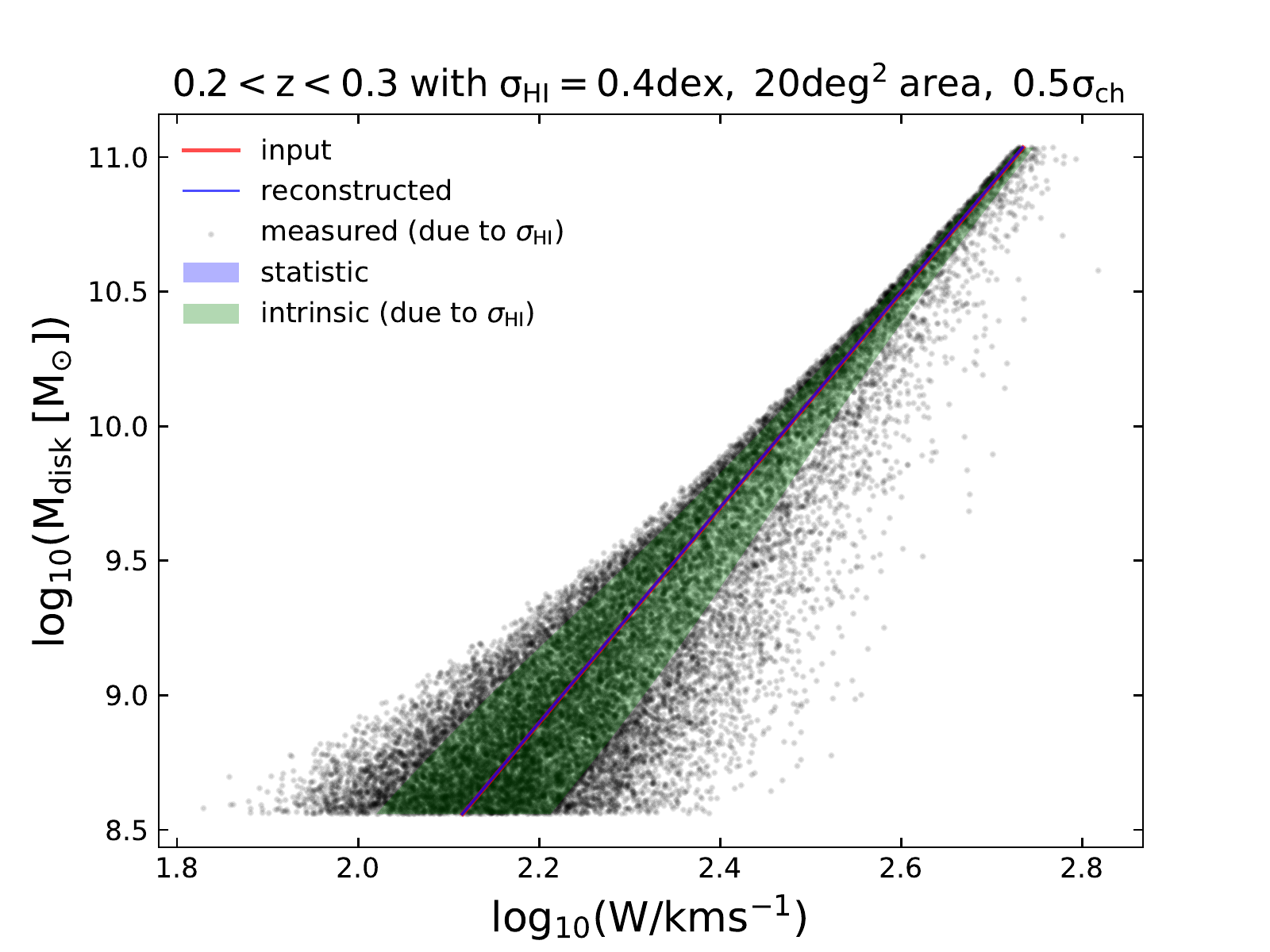}
  \end{subfigure}%
  \hfill
  \begin{subfigure}[b]{0.33\textwidth}
    \includegraphics[width=1.1\columnwidth, height=0.86\columnwidth]{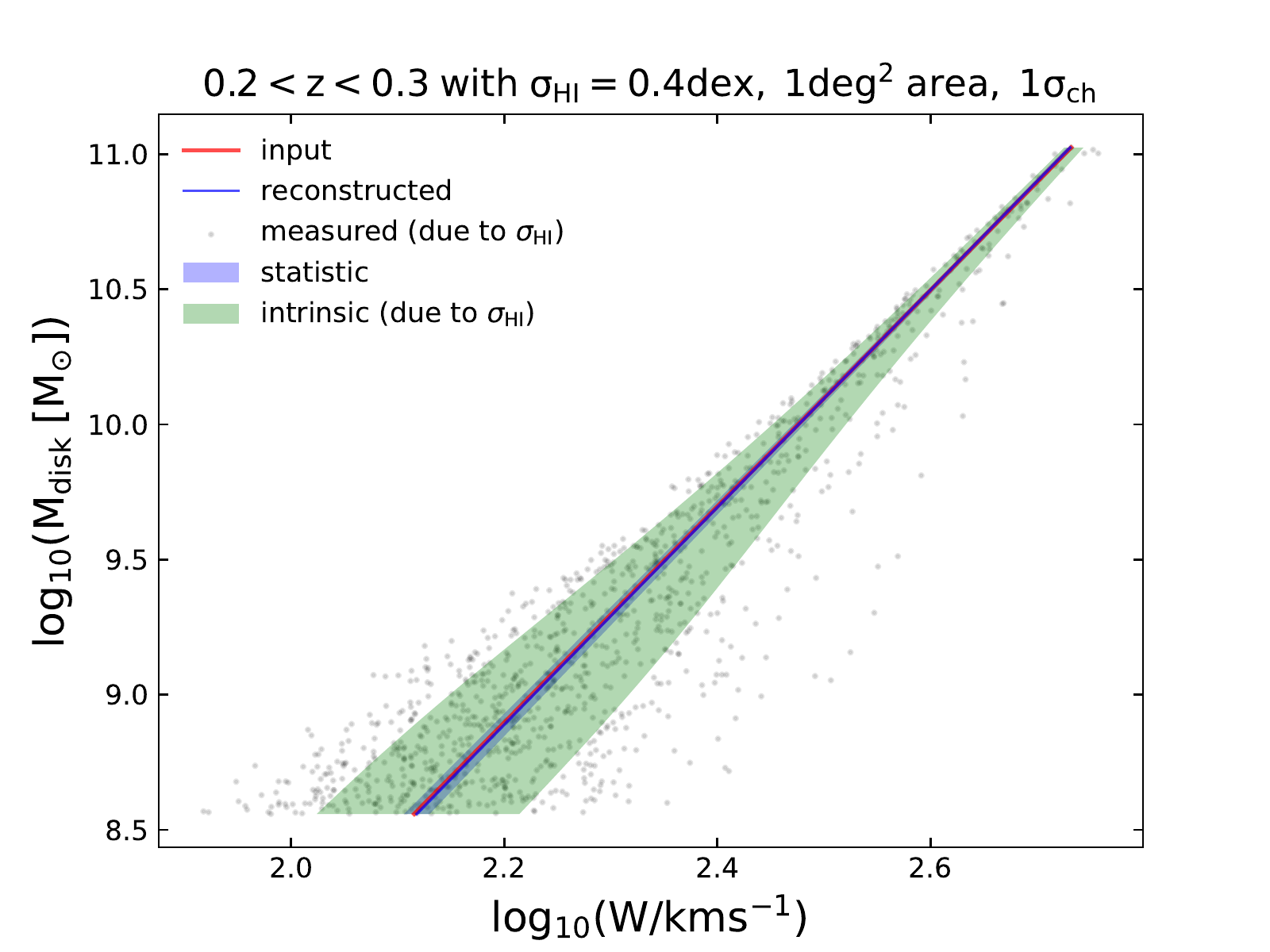}
    \includegraphics[width=1.1\columnwidth, height=0.86\columnwidth]{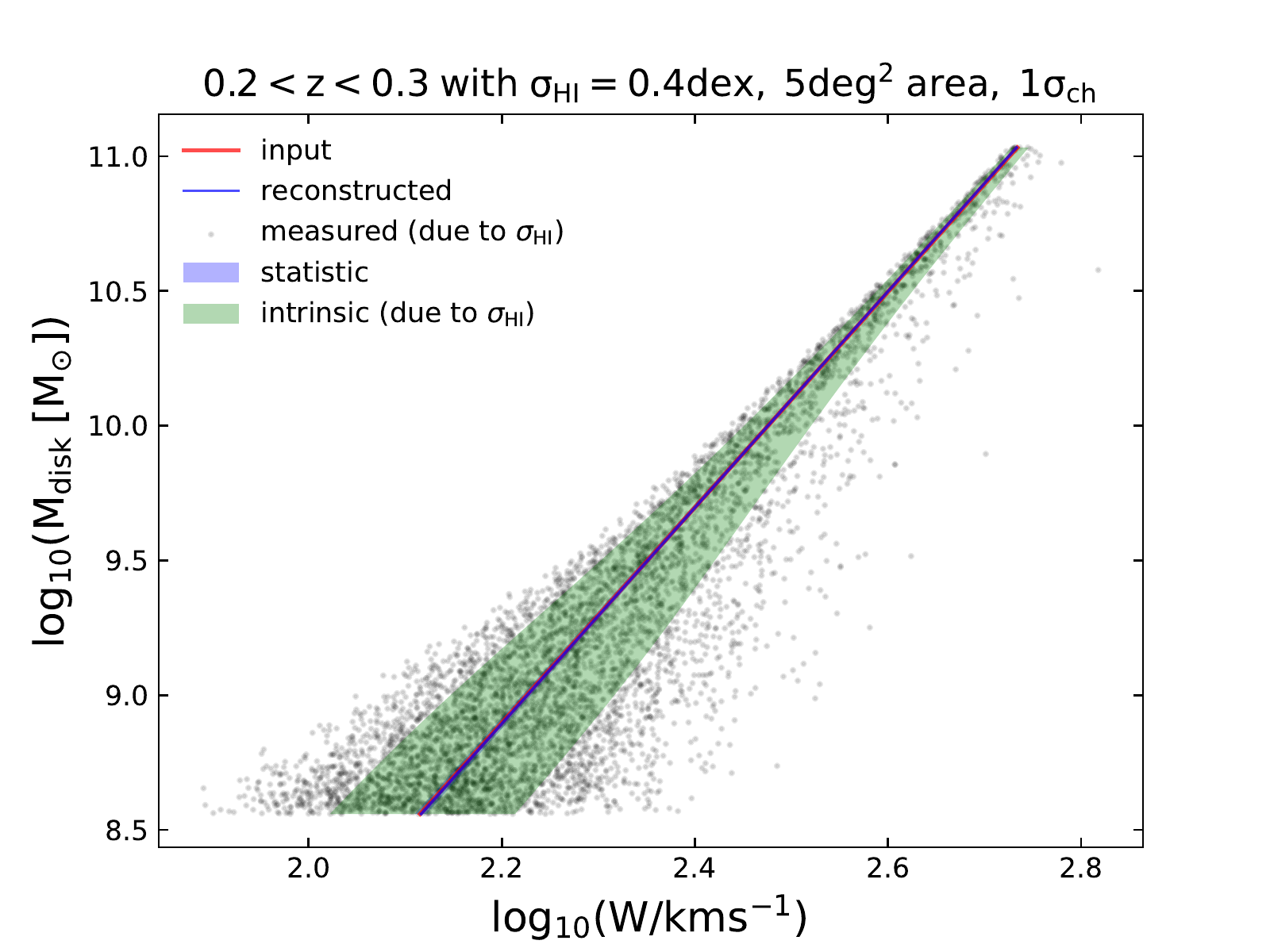}
    \includegraphics[width=1.1\columnwidth, height=0.86\columnwidth]{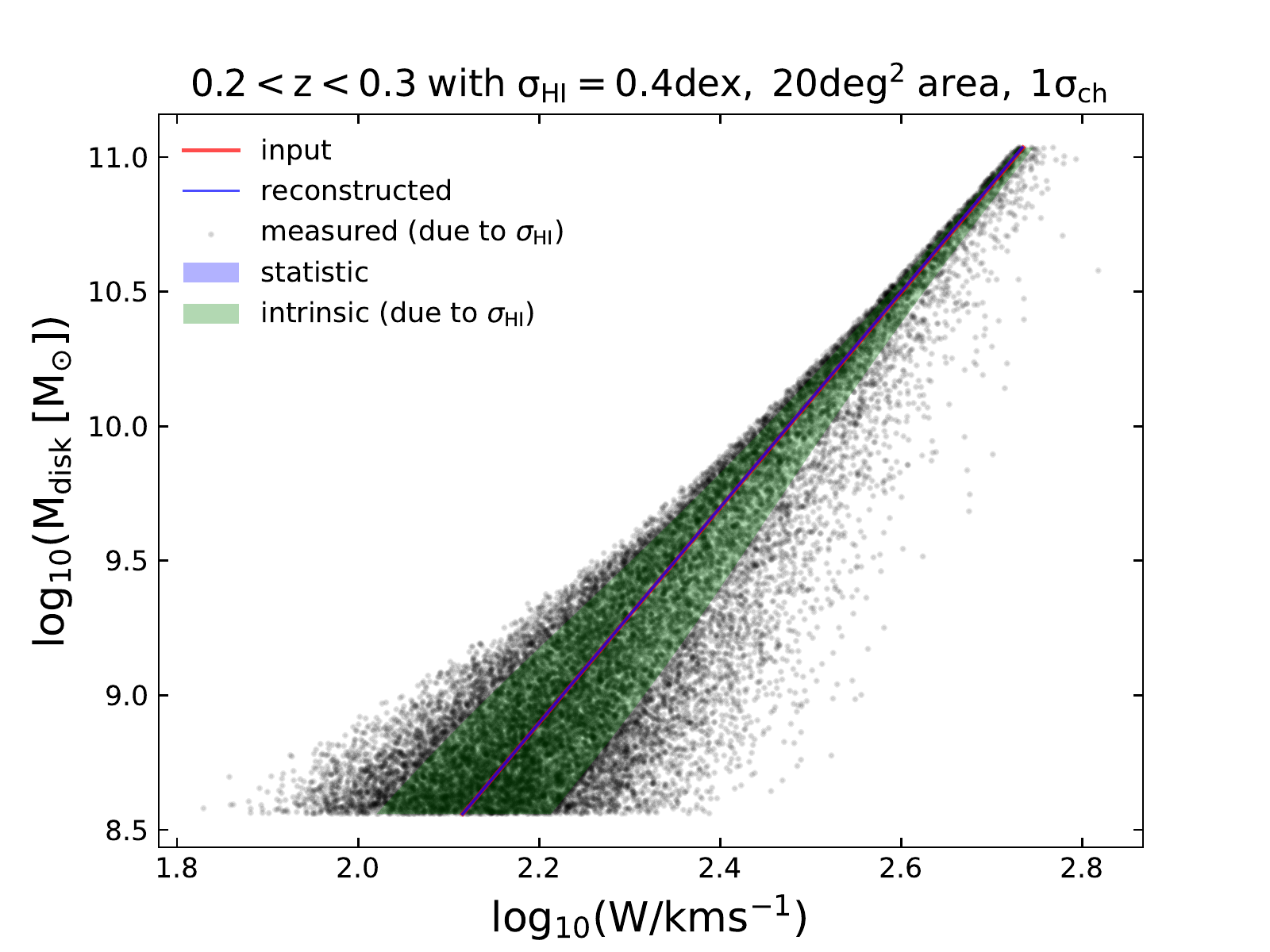}
  \end{subfigure}%
  \hfill
  \begin{subfigure}[b]{0.33\textwidth}
    \includegraphics[width=1.1\columnwidth, height=0.86\columnwidth]{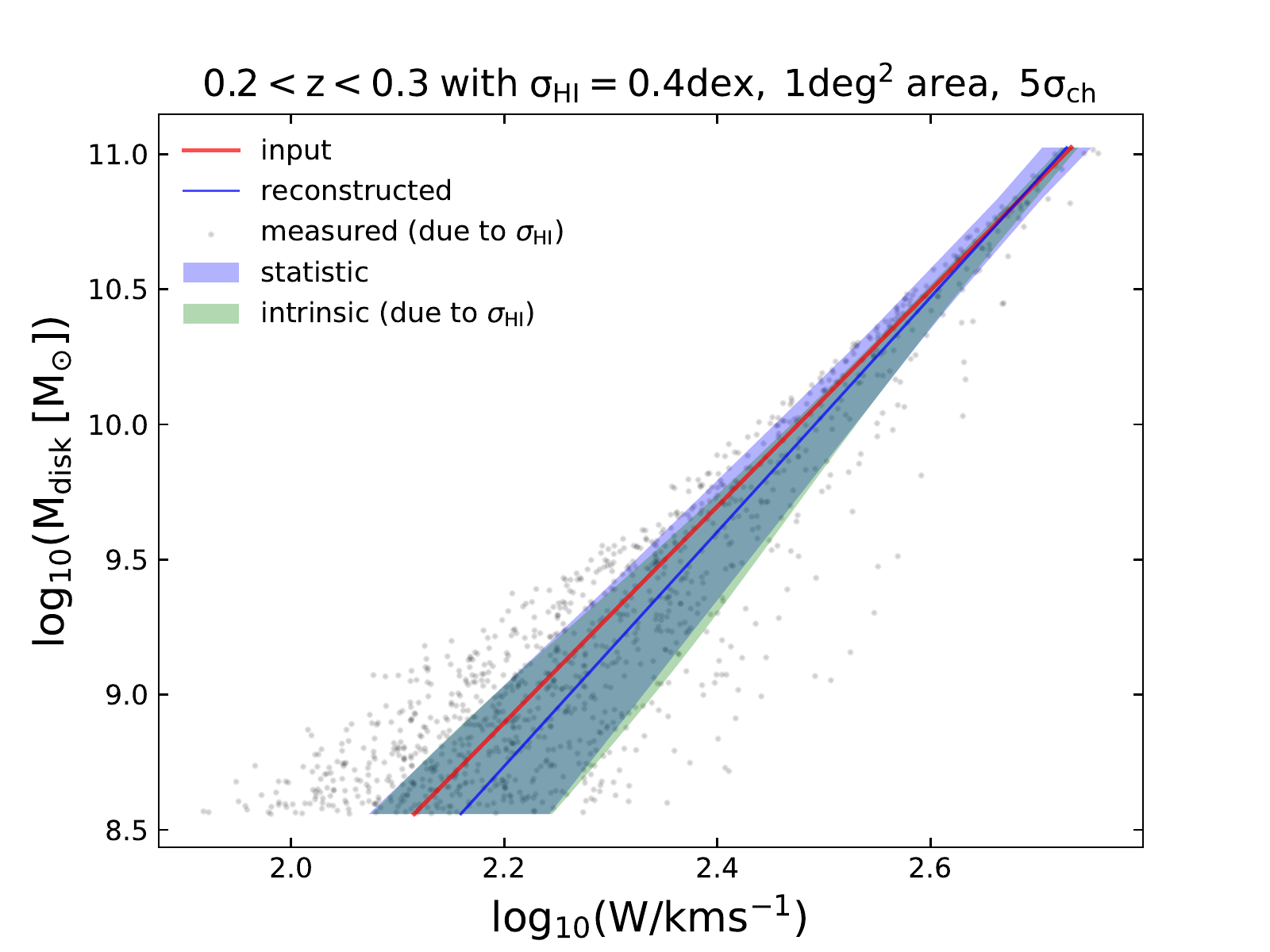}
    \includegraphics[width=1.1\columnwidth, height=0.86\columnwidth]{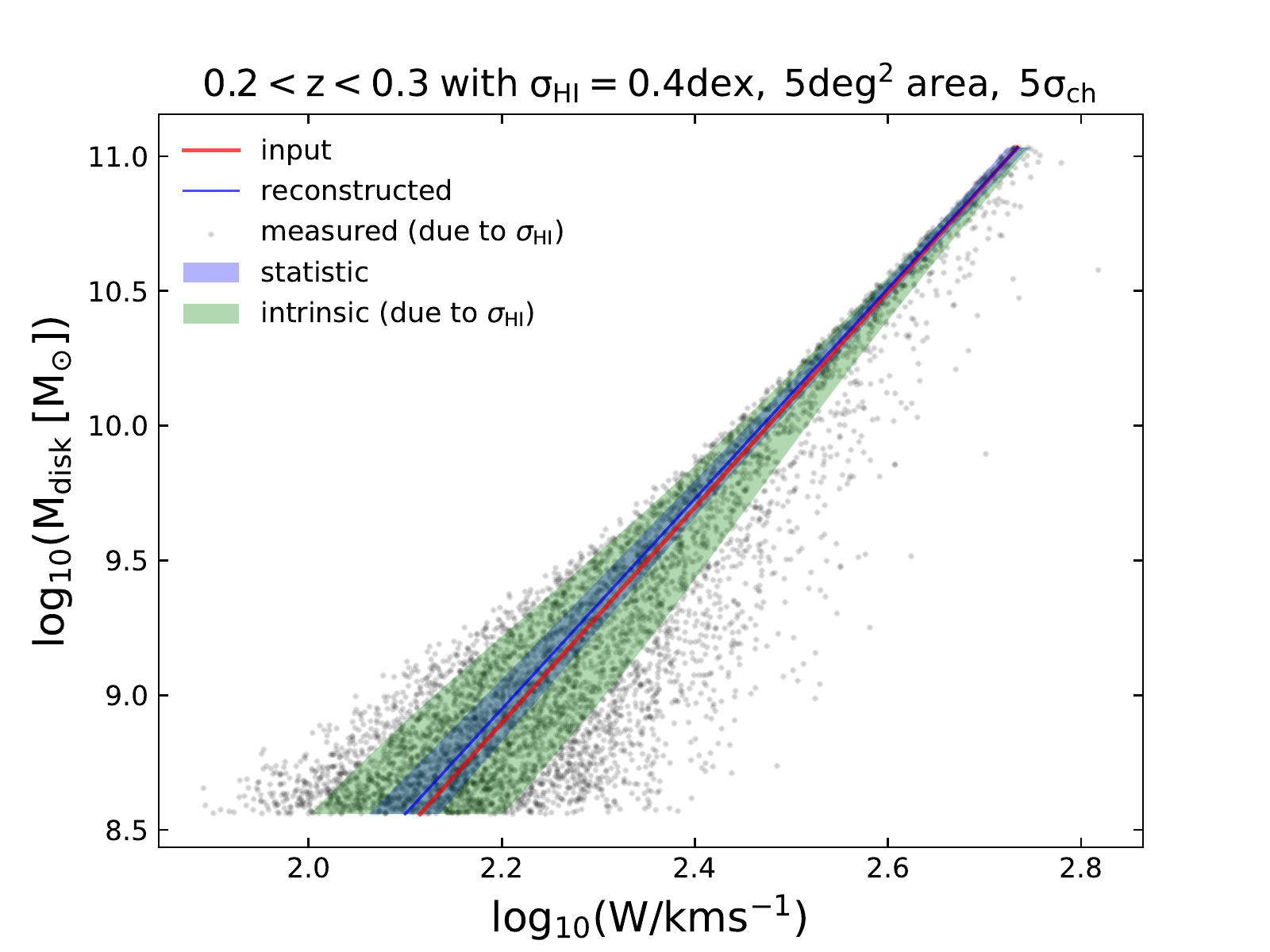}
    \includegraphics[width=1.1\columnwidth, height=0.86\columnwidth]{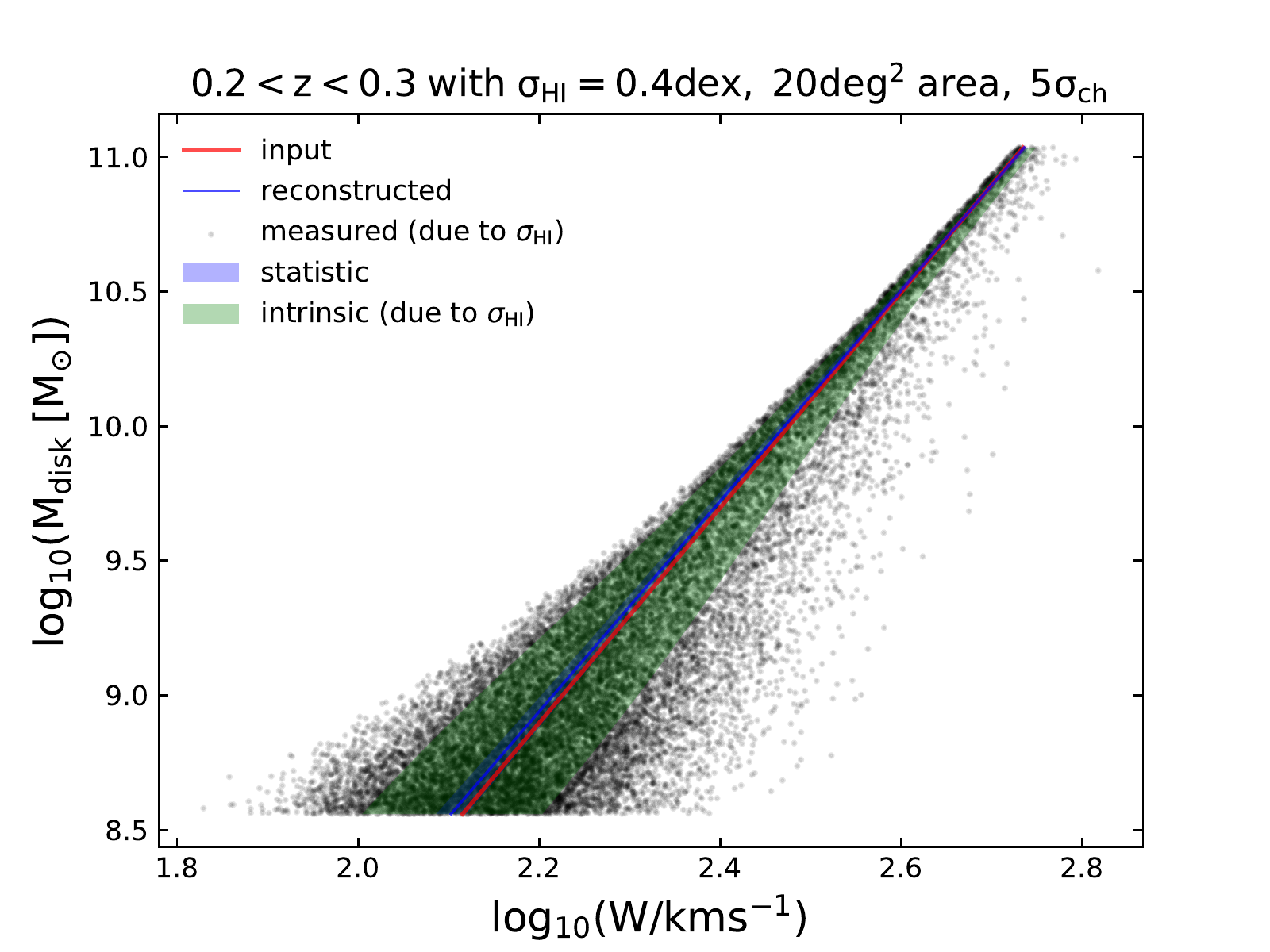}
  \end{subfigure}%
 \caption{Reconstructed baryonic Tully-Fisher relation from a single z-bin ($0.2 < z < 0.3$) with the survey area increased from 1 to 20\,deg$^{2}$ from top to bottom panels. Left: reconstructed from 0.5$\sigma_{\rm ch}$ noise. Middle: 1$\sigma_{\rm ch}$ noise. Right: 5$\sigma_{\rm ch}$ noise.}\label{fig:mdisk-w_160}
\end{figure*}

\subsection{The likelihood function}
\label{sec:Likelihood} 

Based on  Bayes' theorem, the probability of the model given the data is proportional to the likelihood function (i.e. the probability of the data given the model). With the assumption that the behaviour of the noise on intrinsic flux densities of different sources is independent, the likelihood for all the sources having the measured flux densities with our model in a \ha survey is given by

\begin{equation}
    \mathcal{L} \propto  \prod_{\rm source} P(S_{\rm m}(v)|A, \alpha, \sigma_{\rm HI}, b, r_{\rm c}), 
	\label{eq:likeli}
\end{equation}
where $P(S_{\rm m}(v)|A, \alpha, \sigma_{\rm HI}, b, r_{\rm c})$ is the probability of having the measured flux density $S_{\rm m}(v)$ for a single source from Eq.~\eqref{eq:probability}, given the described 5-parameter model above. 


As the PDF for the \ha mass is characterised in the logarithmic space, we rewrite the Eq.~\eqref{eq:probability} as
\begin{multline}
P(S_{\rm m}(v)|A, \alpha, \sigma_{\rm HI}, b, r_{\rm c}) \\
=  \int {\rm d}\log_{10}(M_{\rm HI})P(\log_{10}(M_{\rm HI})) \prod_{v} P_n(S_{\rm m}(v)-S(v, M_{\rm HI})),
\label{eq:probability_log}
\end{multline}
where $P(\log_{10}(M_{\rm HI}))$ follows a normal distribution with mean \ha mass determined by Eq.~\eqref{eq:HI-star} and standard deviation $\sigma_{\rm HI}$. The upper and lower limits of the integral are set to be the mean \ha mass $\pm 2$ dex.

At $z < 0.1$, we marginalise the probability over the extra parameters $b$ and $r_{\rm c}$ to suppress the effect of low-number statistics as
\begin{multline}
P(S_{\rm m}(v)|A, \alpha, \sigma_{\rm HI}) \\
= \int d\log_{10}(M_{\rm HI})P(\log_{10}(M_{\rm HI})) \int dbP(b) \int dr_{\rm c}P(r_{\rm c}) \\ \prod_{v} P_n(S_{\rm m}(v)-S(v, M_{\rm HI})),
\label{eq:probability_marginal}
\end{multline}
where $P(b)$ and $P(r_{\rm c})$ follow uniform distributions as listed in Table~\ref{tab:priors}.

In practice, we take the logarithm of the likelihood function:
\begin{equation}
 \ln \mathcal{L} = \sum_{\rm source} \log(P(S_{\rm m}(v))) + \rm constant.
\label{eq:logp_z}
\end{equation}

\subsection{Parameter estimation}

 We use {\sc Multinest} \citep{feroz2009multinest, buchner2014x} to sample the prior parameter space. {\sc Multinest} is an efficient and robust tool for the Bayesian inference, which produces the posterior parameter samples with an associated error estimate. Priors capture our knowledge of a parameter before any new experimental evidence is taken into account. These are listed in Table~\ref{tab:priors}.

\begin{table}
	\centering
	\caption{The input parameters of the model for measuring the Tully-Fisher relation, \ha and stellar mass relation and double-horn profiles.}
	\label{tab:priors}
	\begin{tabular}{lrl}
	\hline
	Parameter & Input & Prior Probability Distribution \\
	\hline
	$\log_{10}(A)$     & 9.3         & uniform $\in [7, 12] $       \\
	$\alpha $          & 4           & uniform $\in [2.5, 7.5] $    \\
	\hline
	$\sigma_{\rm HI}$          & 0.4           & uniform $\in [0.1, 0.9] $    \\
    \hline
    $b$          &  [0.3, 0.7]           & uniform $\in [0, 1] $    \\
    $r_{\rm c}$          & [1.5, 2]           & uniform $\in [1, 3] $    \\
    \hline
	\label{tab:priors}
	\end{tabular}
\end{table}

\subsection{Survey parameters}

Following P20, we initially assume a 1\,deg$^{2}$ sky area with noise $\sigma_{\rm ch} = 90$\,$\mu$Jy/channel over the redshift range $0<z<0.55$ as our baseline survey. We then investigate the effect of changing the noise properties of the spectral line cube and the survey area to investigate how they affect our ability to measure the bTFr. Specifically, we change the RMS of the spectral line cubes to $0.5\sigma_{\rm ch}$ and $5\sigma_{\rm ch}$, and increase the survey area from 1\,deg$^{2}$ to 20\,deg$^{2}$. We also simulate a second group of galaxies for surveys with the same parameter setup, but with only Gaussian emission line profiles to investigate the effect of fitting double-horn profiles. To demonstrate the key elements of our work, we incorporate the scatter in the $M_{\rm HI} - M_{\star}$ relationship into our model but fix the slope and normalisation, although we note that these could also be included in the modelling.

Here we set the redshift bin width to be 0.1 for studying the effect of redshift evolution on our approach. This binning scheme is relatively coarse rather than optimal to demonstrate that our approach is insensitive to the way of binning the datasets. In principle, a smaller bin width to further extract hidden information could be used.

\section{Results}
\label{sec:results}

\subsection{The reconstructed Tully-Fisher relation from our baseline survey}

\begin{figure}
 \includegraphics[width=1.0\columnwidth]{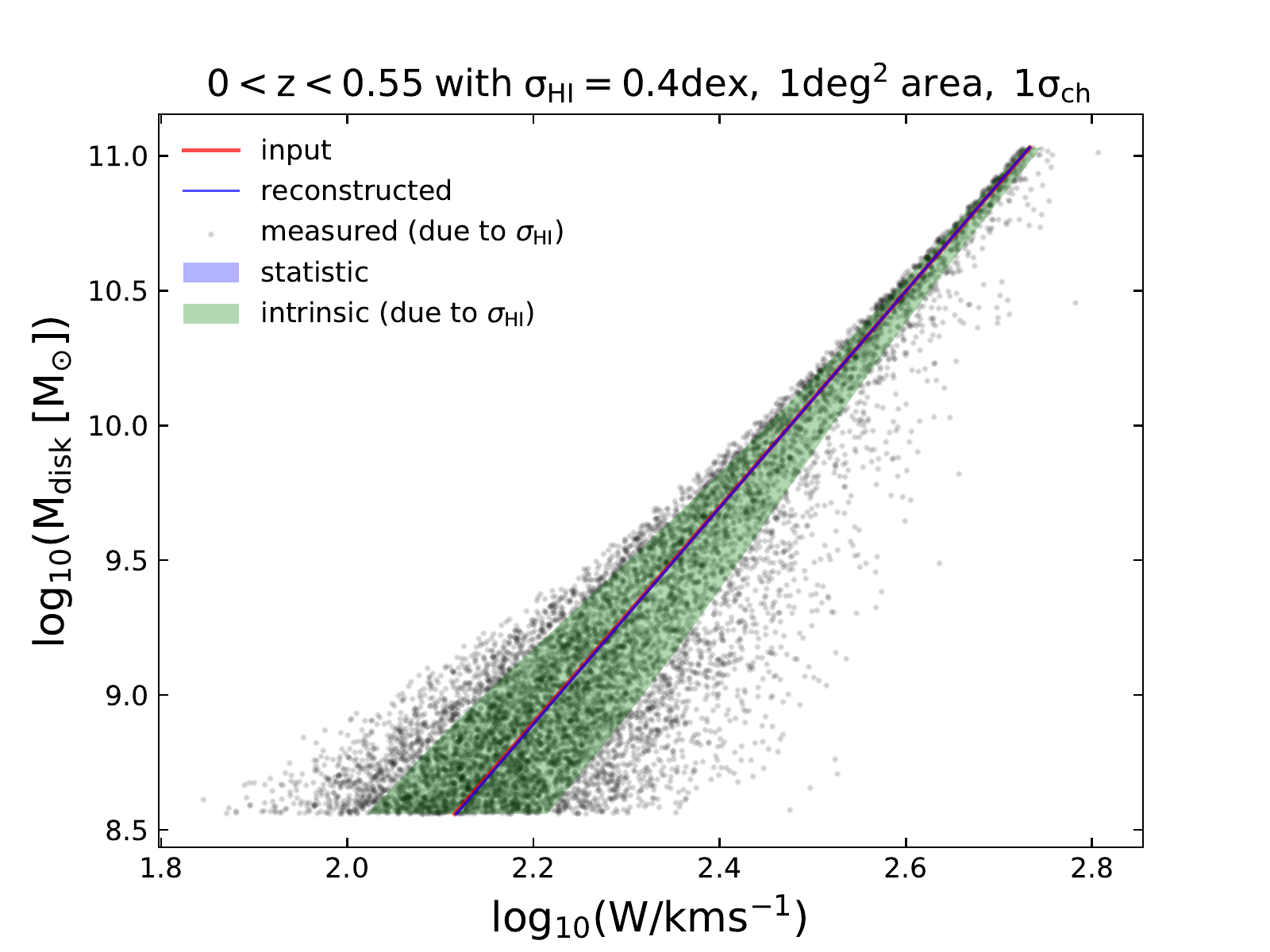}
 \caption{The reconstructed baryonic Tully-Fisher relation from a broad z-bin with a 1$\sigma_{\rm ch}$ noise.}\label{fig:mdisk-w}
\end{figure}

Figure~\ref{fig:mdisk-w_160} shows the reconstructed baryonic Tully-Fisher relation with $\sigma_{\rm HI}$ =0.4 dex on the $M_{\rm HI}-M_{\star}$ relation at $0.2 < z < 0.3$. The result in the top-middle panel is the reconstructed bTFr for our baseline survey. The input (simulated) relation is shown by the red line (in the absence of any scatter). The simulated disk masses $M_{\rm disk}$ are denoted by the black points. We note that these disk masses are estimated using the prior knowledge of the stellar mass and the $M_{\rm HI}-M_{\star}$ relation along with the scatter (without knowing the accurate \ha mass). The velocity widths $W$ on the abscissa, are the input values based on the input \ha mass. Therefore, the scatter is only present in the measured disk masses (grey points). In reality, the scatter may be higher as there is also scatter in the \ha mass to the velocity width for individual galaxies, which would manifest in the scatter on the bTFr itself.

\begin{figure*}
  \centering
  \begin{subfigure}[b]{0.33\textwidth}
    \includegraphics[width=1.\columnwidth, height=1.\columnwidth]{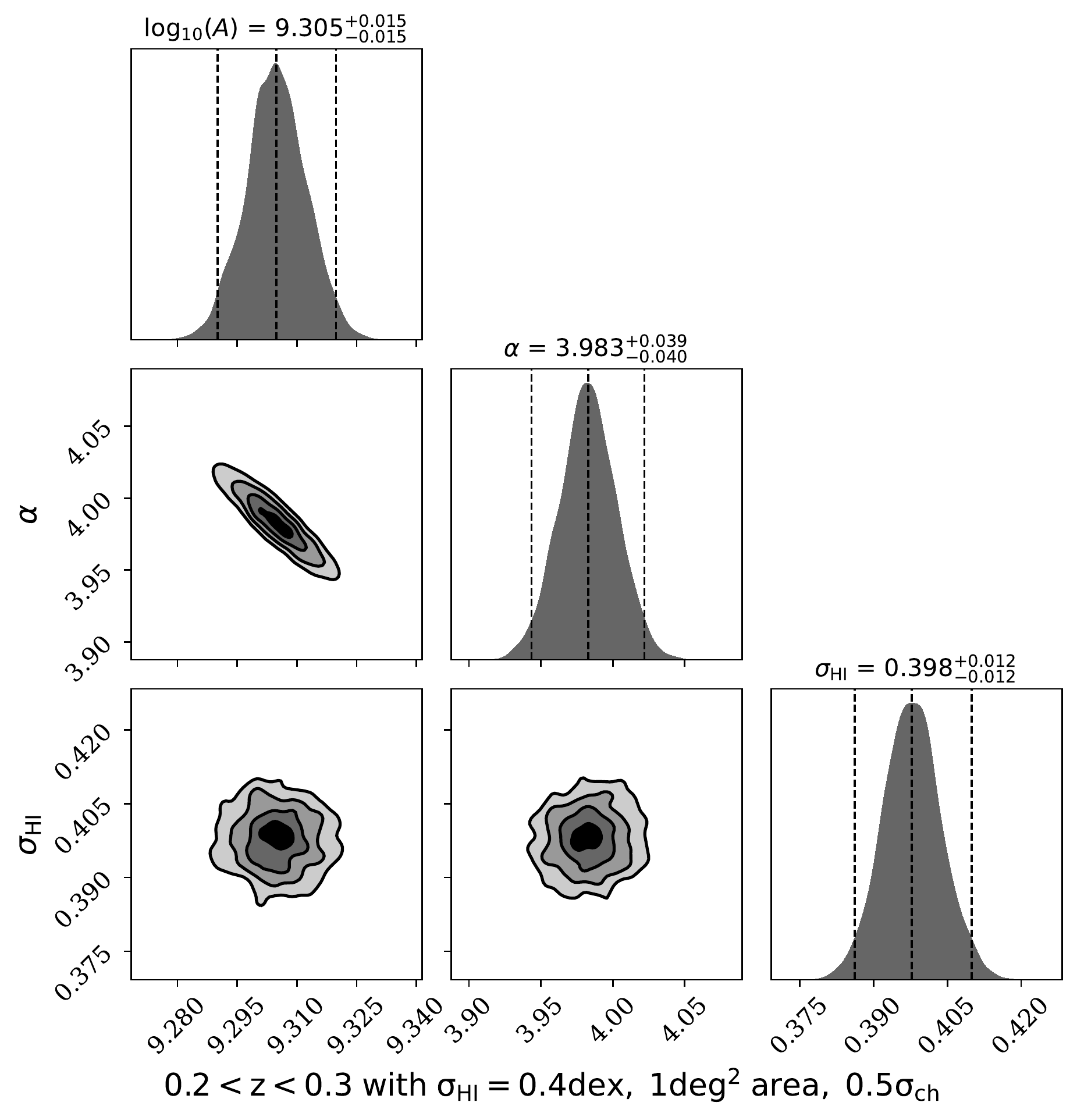}
    \includegraphics[width=1.\columnwidth, height=1.\columnwidth]{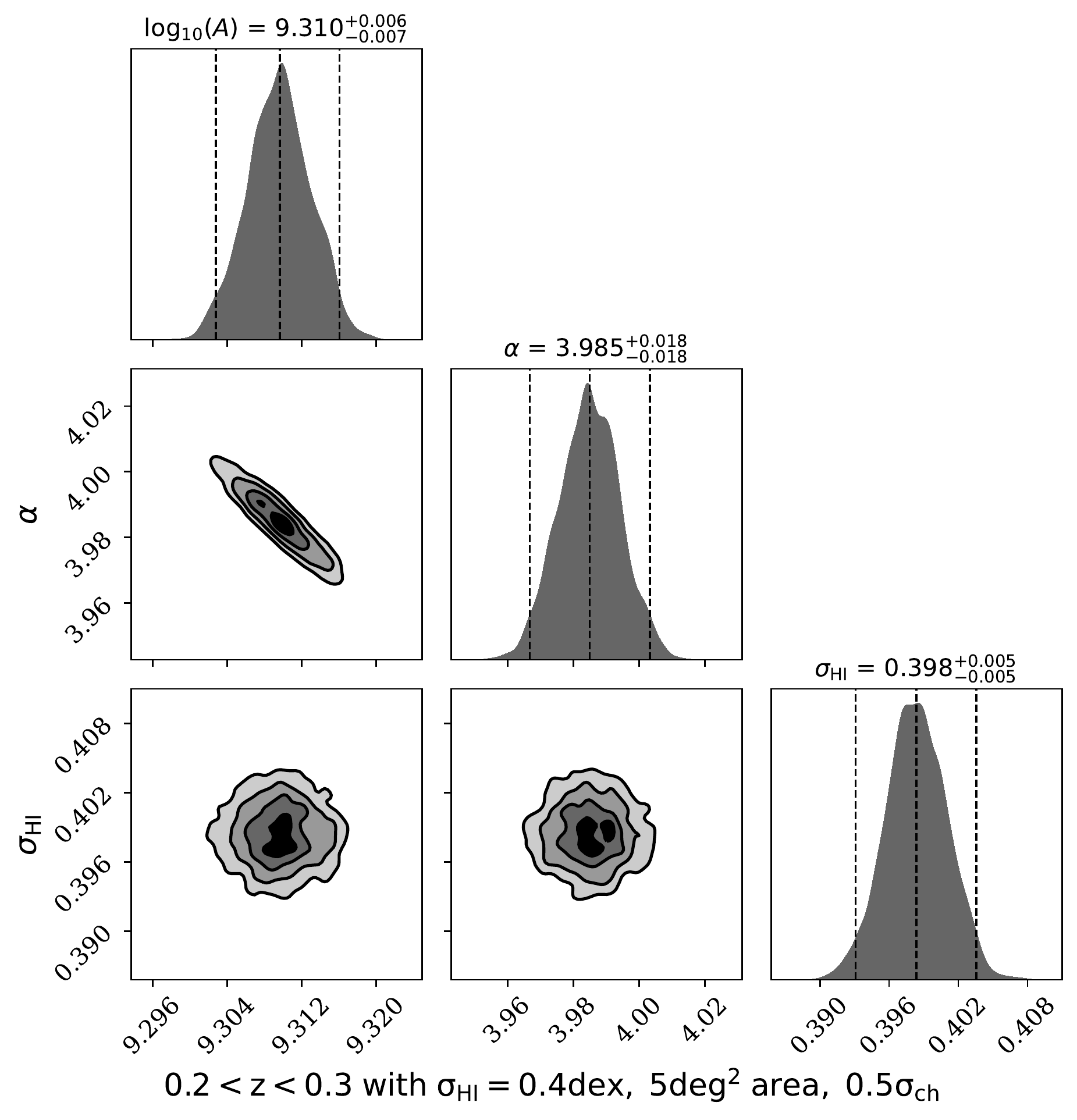}
    \includegraphics[width=1.\columnwidth, height=1.\columnwidth]{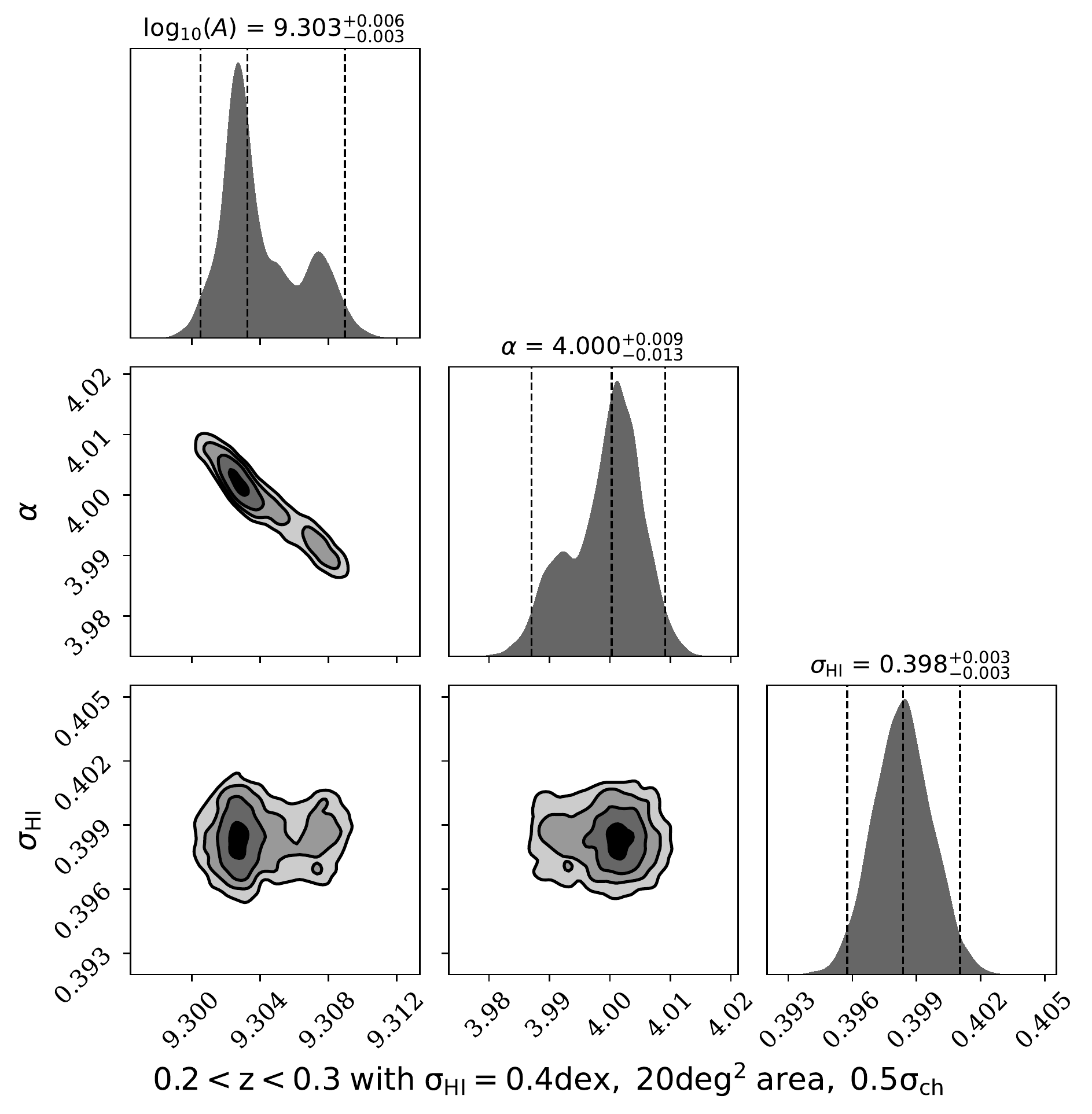}
  \end{subfigure}%
  \hfill
  \begin{subfigure}[b]{0.33\textwidth}
    \includegraphics[width=1.\columnwidth, height=1.\columnwidth]{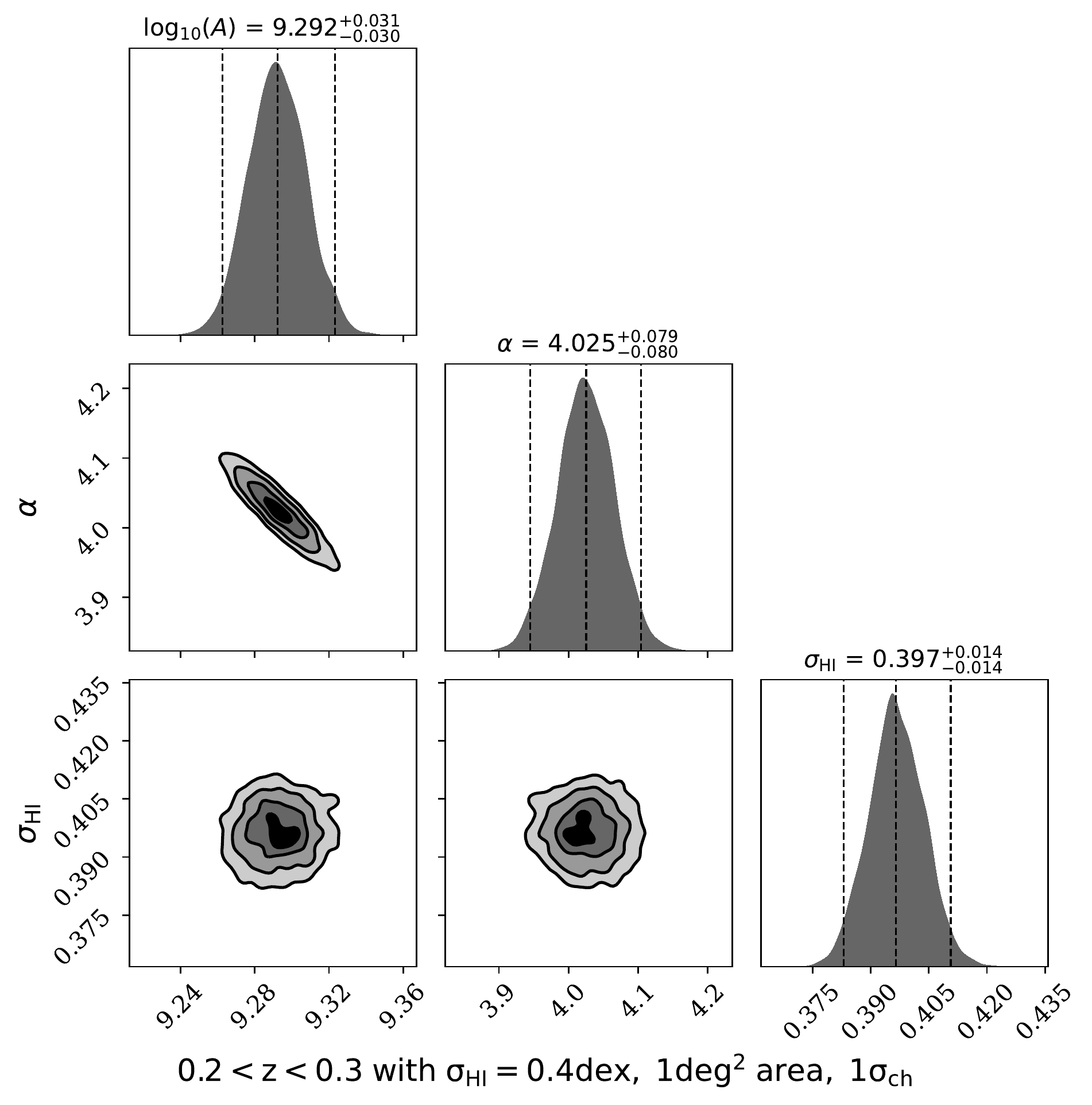}
    \includegraphics[width=1.\columnwidth, height=1.\columnwidth]{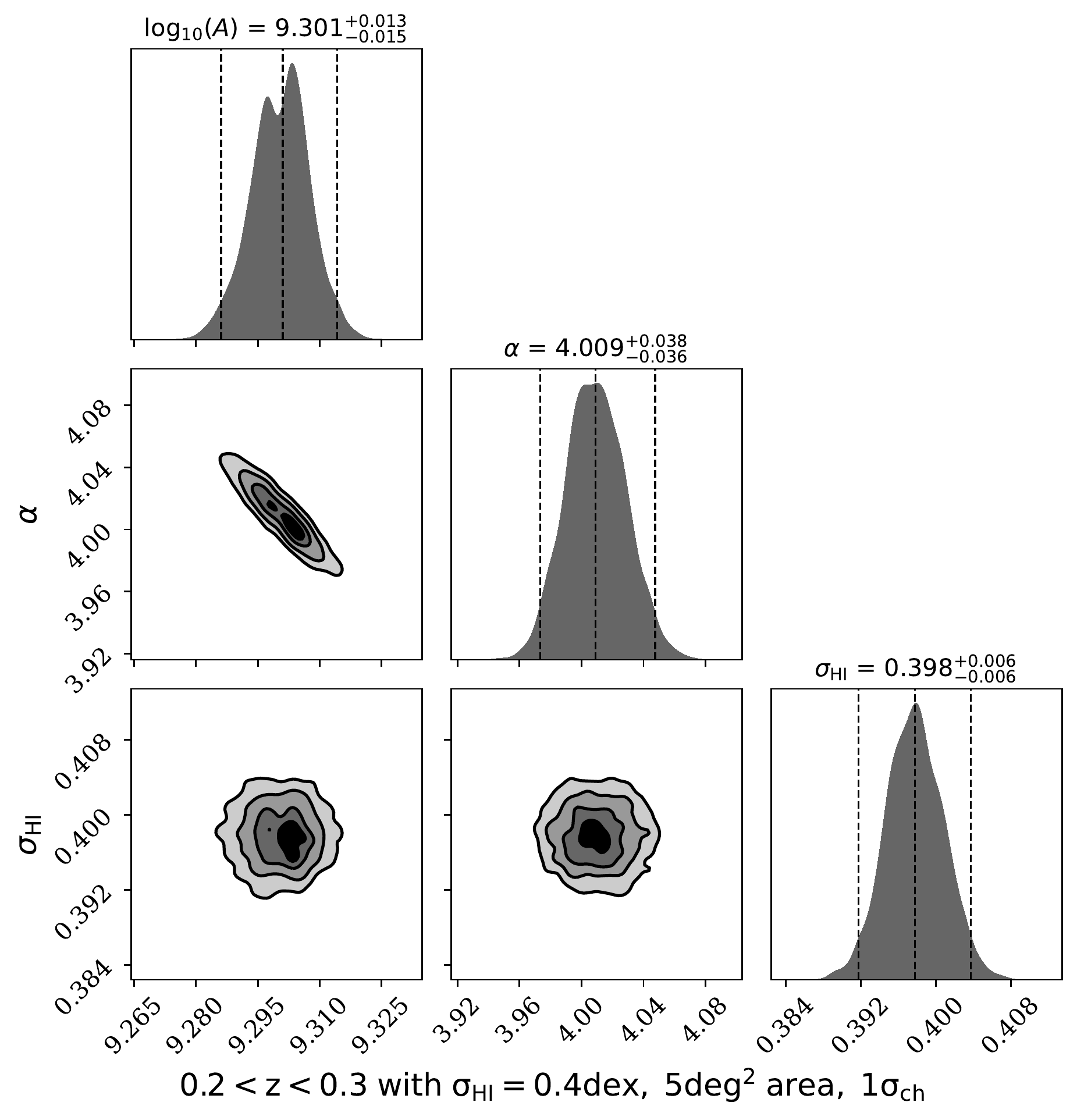}
    \includegraphics[width=1.\columnwidth, height=1.\columnwidth]{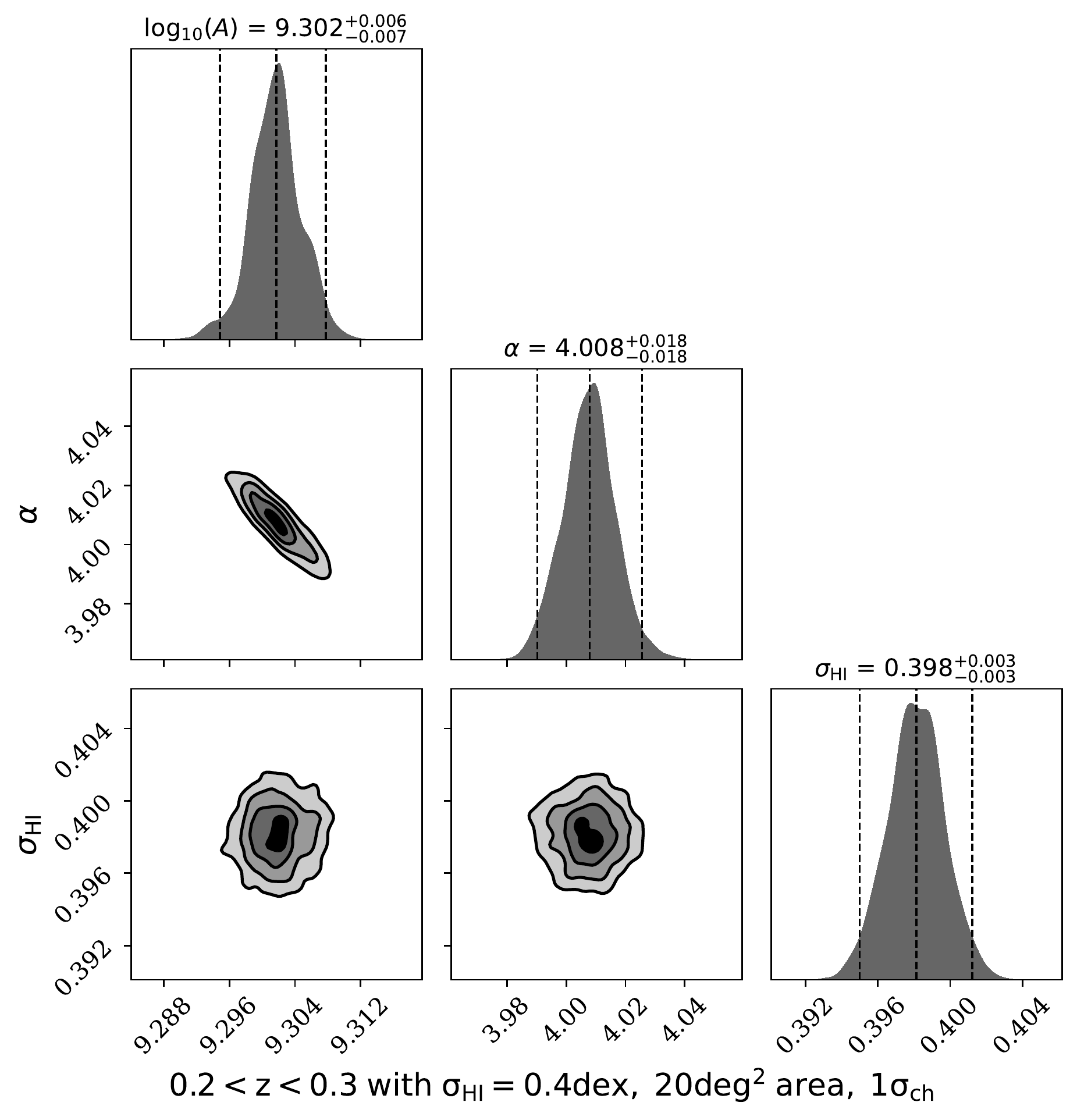}
  \end{subfigure}%
  \hfill
  \begin{subfigure}[b]{0.33\textwidth}
    \includegraphics[width=1.\columnwidth, height=1.\columnwidth]{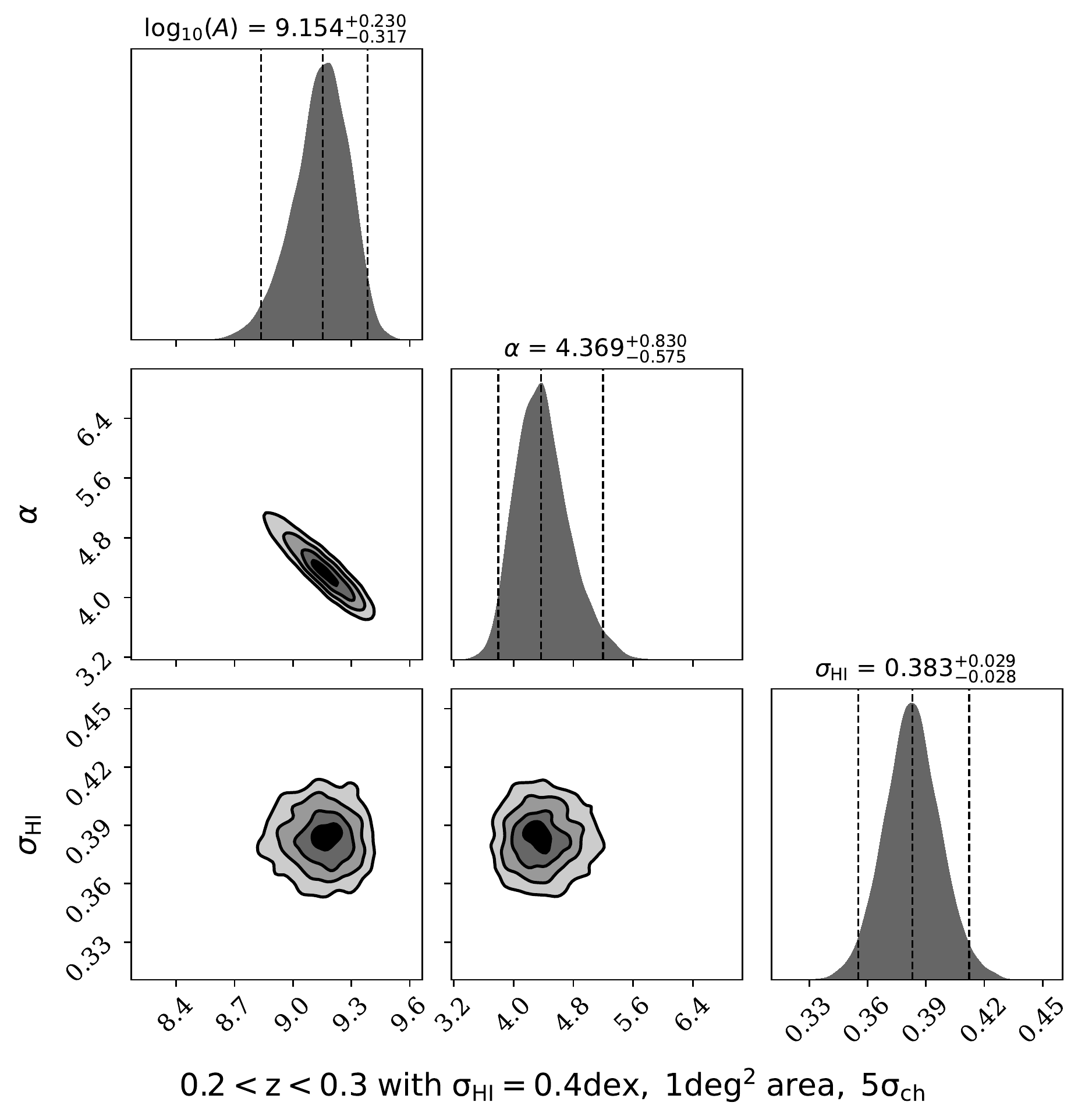}
    \includegraphics[width=1.\columnwidth, height=1.\columnwidth]{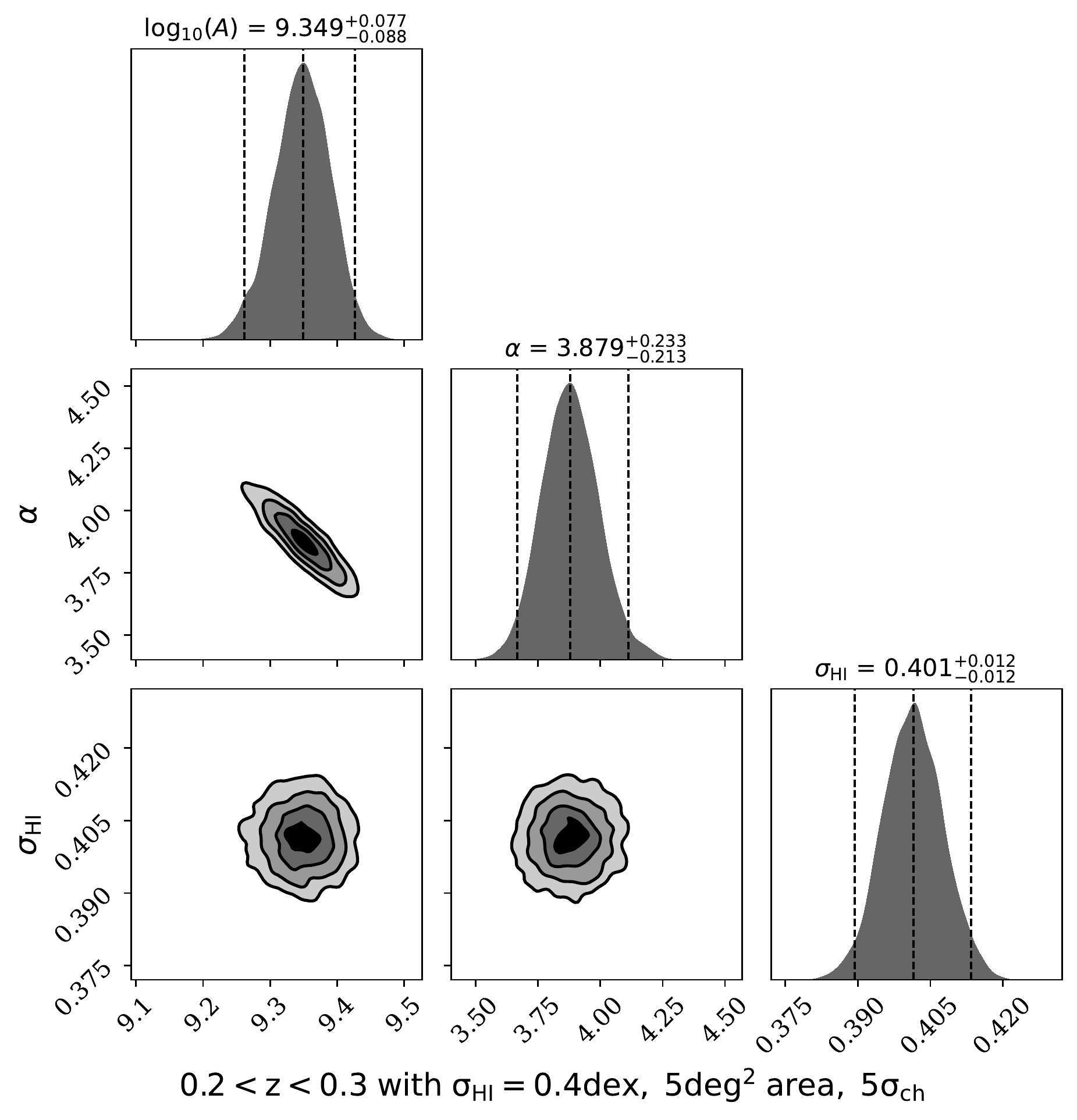}
    \includegraphics[width=1.\columnwidth, height=1.\columnwidth]{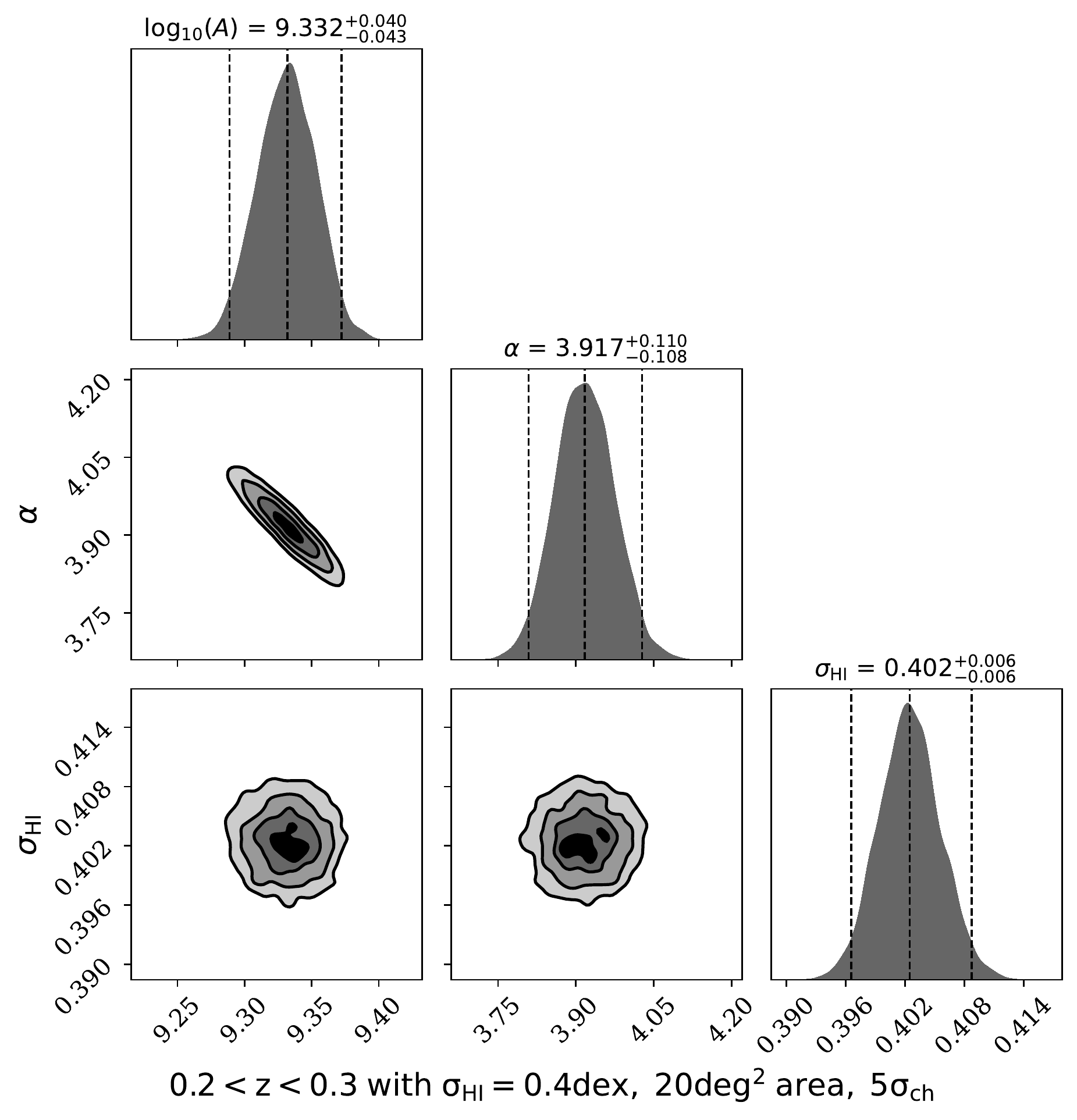}
  \end{subfigure}%
 \caption{The posterior distributions of reconstructed parameters from a single z-bin ($0.2 < z < 0.3$) with the survey area increased from from 1 to 5\,deg$^{2}$ and up to 20\,deg$^{2}$ from top to bottom panels. Note that the scales of axes are different across the panels. Left: reconstructed from 0.5$\sigma_{\rm ch}$ noise. Middle: 1$\sigma_{\rm ch}$ noise. Right: 5$\sigma_{\rm ch}$ noise.}\label{fig:corner_160}
\end{figure*}

The best-fit model of the bTFr using our simulation is shown with the blue line, and the 68\% credible intervals estimated from the posterior distributions of reconstructed parameters $A$ and $\alpha$ (Figure~\ref{fig:corner_160}) are denoted by the blue region. The \ha signal for those galaxies with low disk mass is very faint, and more susceptible to the influence of the noise. Thus, the statistical uncertainty from the measurements becomes larger at low mass. On the other hand, the noise becomes less influential for the more massive galaxies. However, since the number of most massive galaxies is low, the strongest constraints on the bTFr are provided by the middle upper range of disk mass at $M_{\rm disk}\sim10^{10.5}$\,M$_\odot$. This effect can be seen in the top right panel of Figure~\ref{fig:mdisk-w_160}, where we show the constraints on the bTFr for a significantly elevated noise. Therefore, a different form of SMF or colour selection that  only affects the number of objects at the highest mass end will have little impact on our results. 

The intrinsic uncertainty of the bTFr, propagated from the reconstructed $\sigma_{\rm HI}$ on the $M_{\rm HI}-M_{\star}$relation, are shown with the green area. Low-mass galaxies tend to be more \ha dominated, therefore the constant $\sigma_{\rm HI}$ will have a stronger impact on the lower mass end compared to the higher mass end, which tend to be dominated by stellar mass.

\begin{figure*}
    \includegraphics[width=0.85\textwidth]{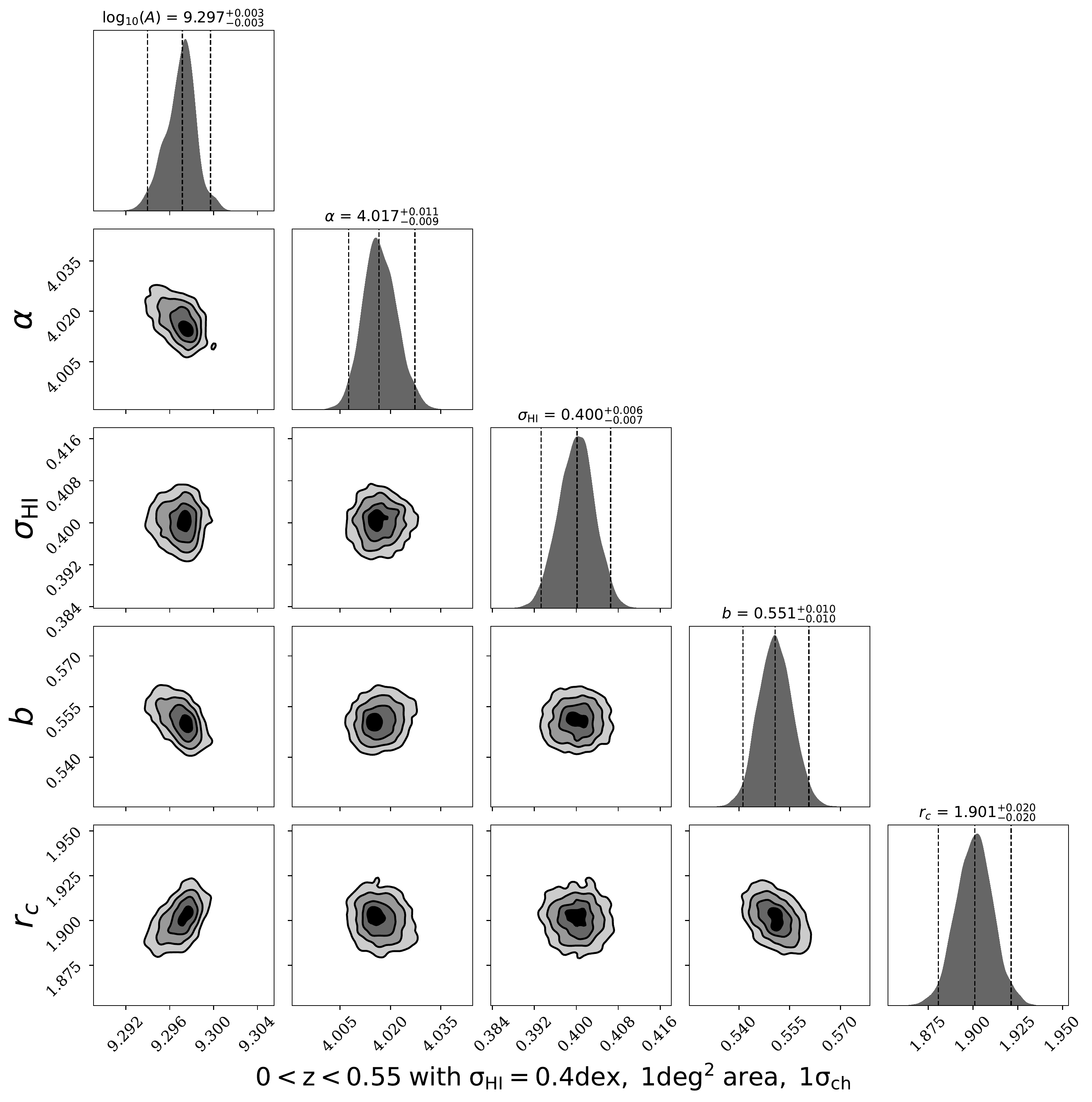}
    \label{fig:sub2}
 \caption{The posterior distributions of reconstructed parameters from a broad z-bin with a 1$\sigma_{\rm ch}$ noise.}\label{fig:corner}
\end{figure*}

\begin{figure*}
  \centering
  \begin{subfigure}[b]{0.33\textwidth}
    \includegraphics[width=1.1\columnwidth, height=0.86\columnwidth]{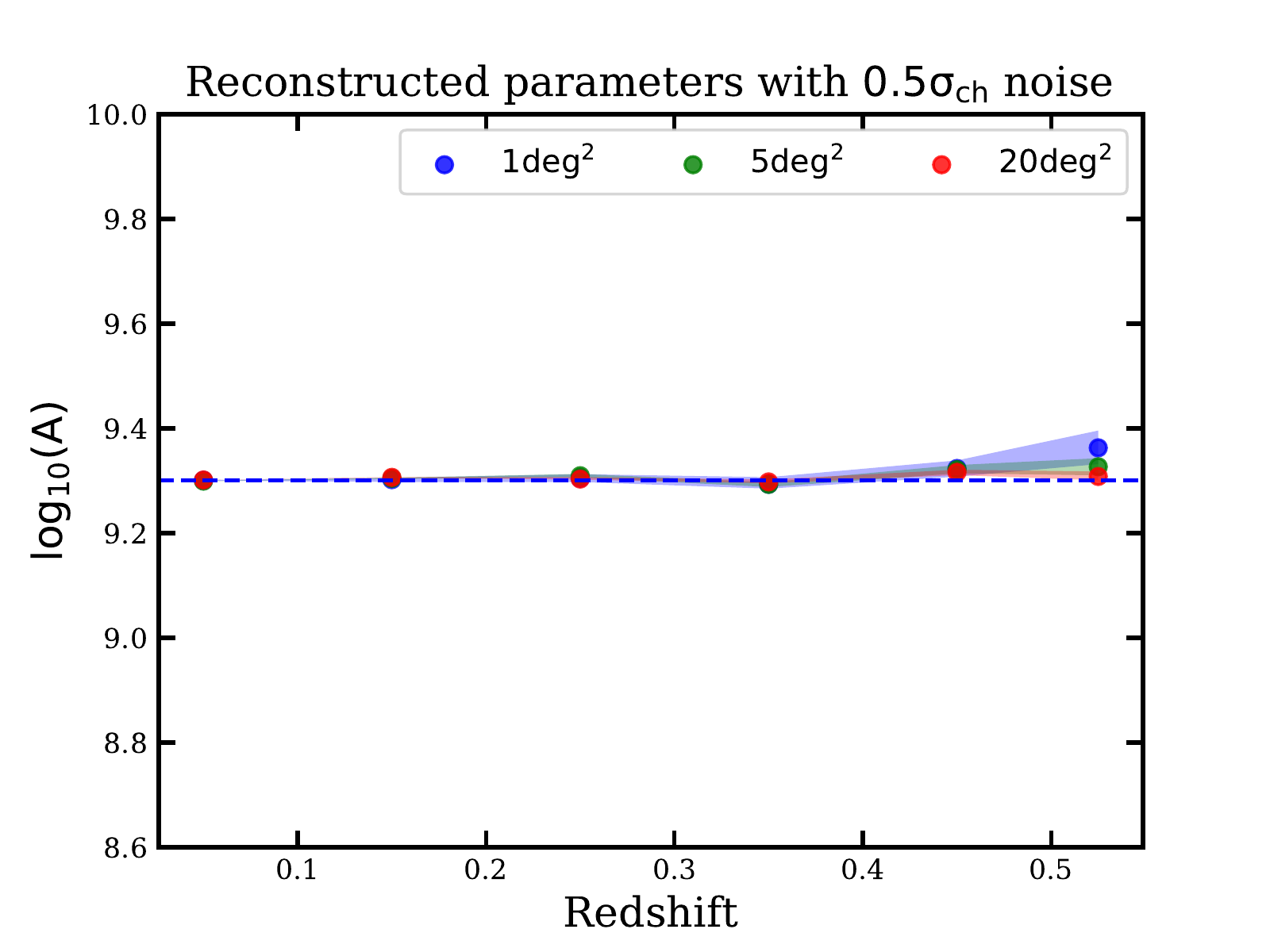}
    \includegraphics[width=1.1\columnwidth, height=0.86\columnwidth]{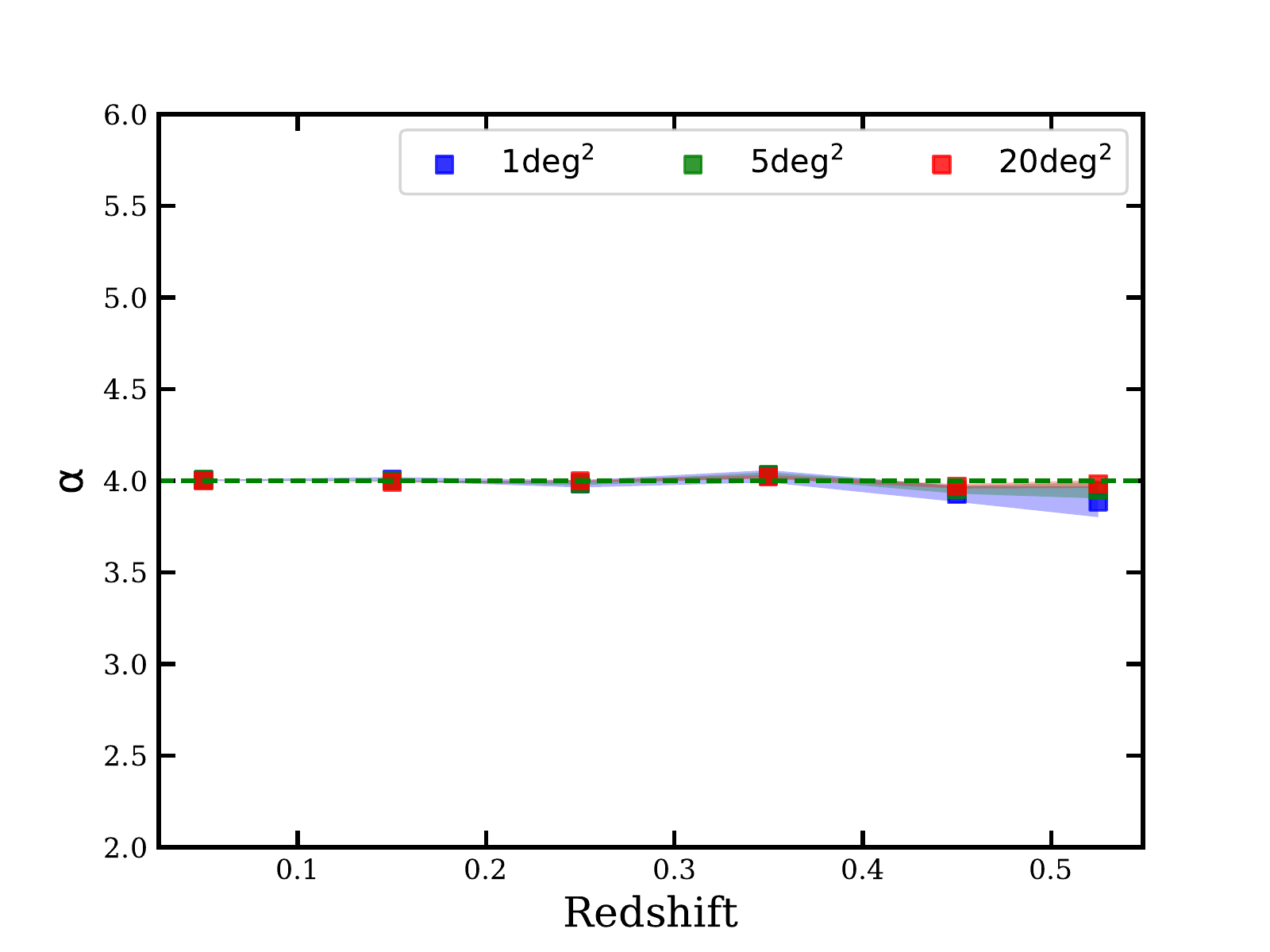}
    \includegraphics[width=1.1\columnwidth, height=0.86\columnwidth]{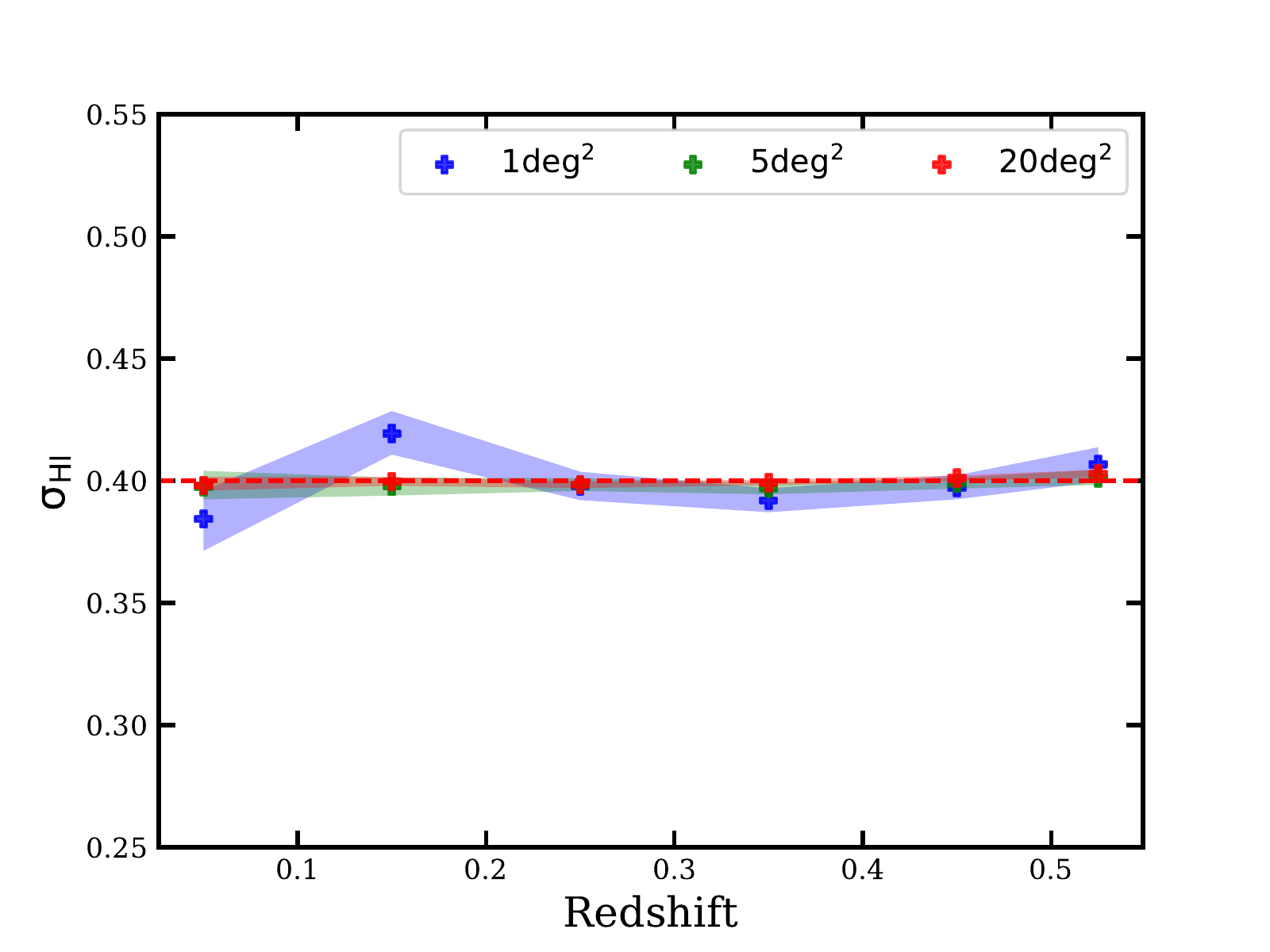}
  \end{subfigure}%
  \hfill
  \begin{subfigure}[b]{0.33\textwidth}
    \includegraphics[width=1.1\columnwidth, height=0.86\columnwidth]{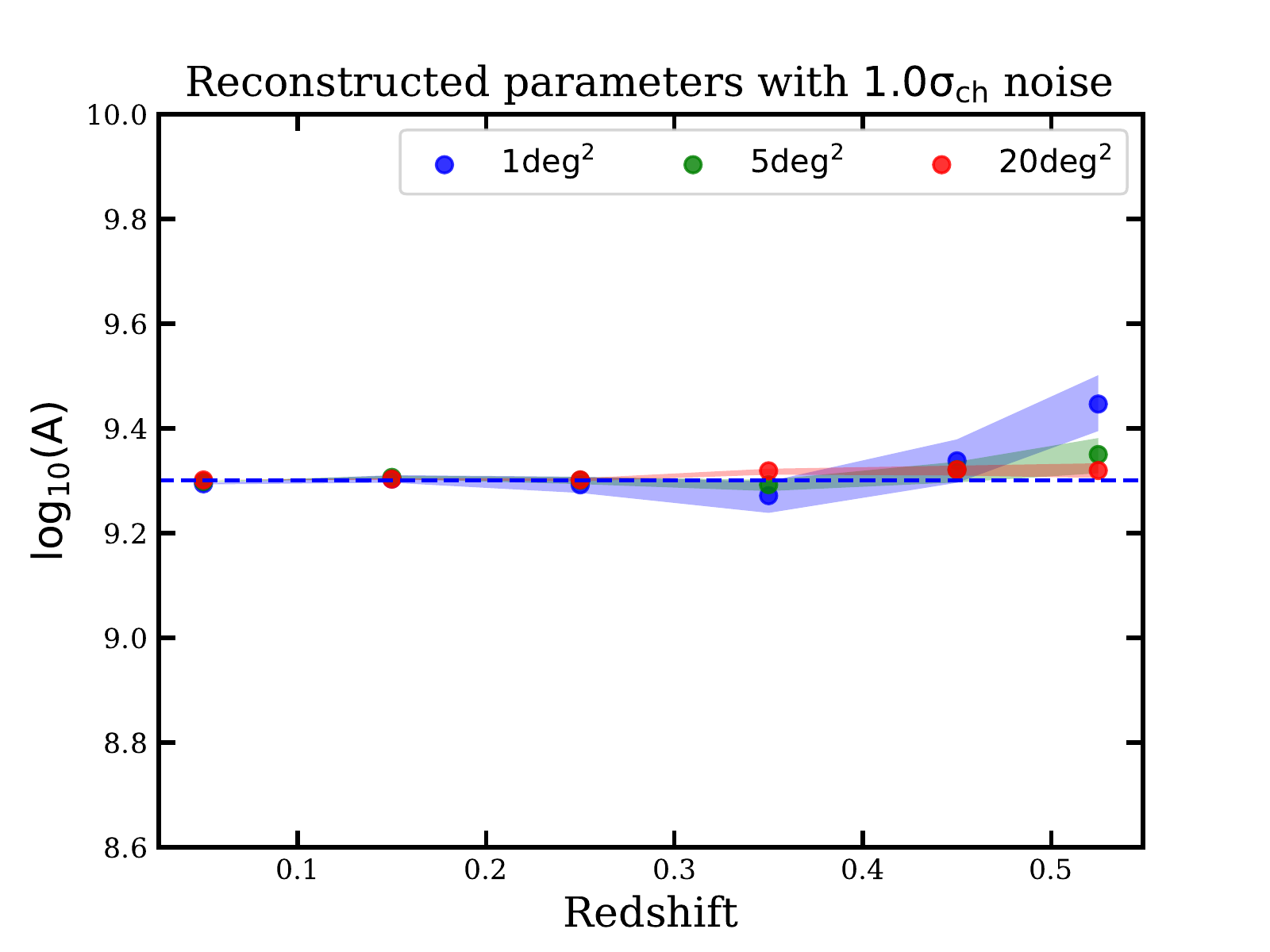}
    \includegraphics[width=1.1\columnwidth, height=0.86\columnwidth]{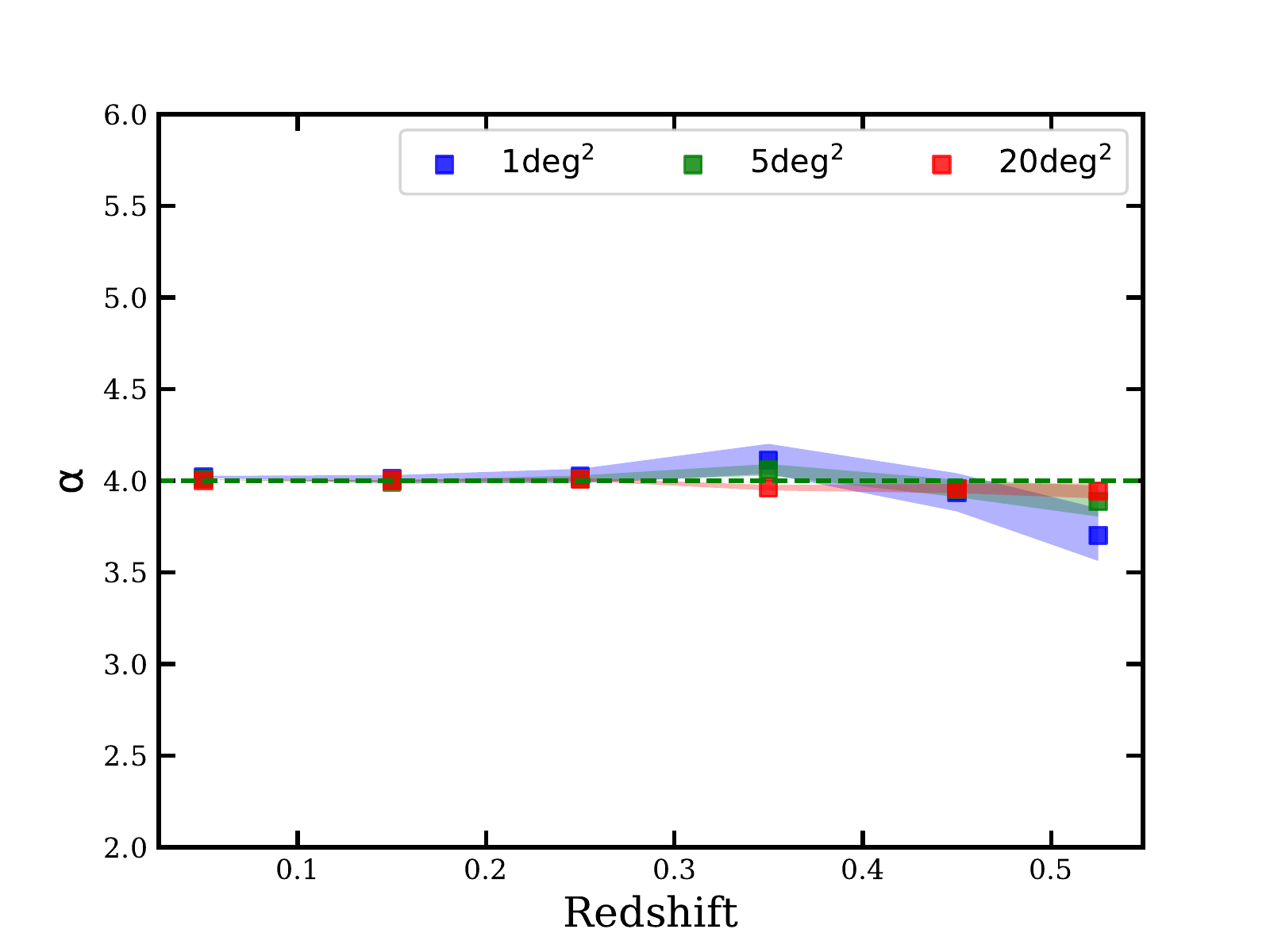}
    \includegraphics[width=1.1\columnwidth, height=0.86\columnwidth]{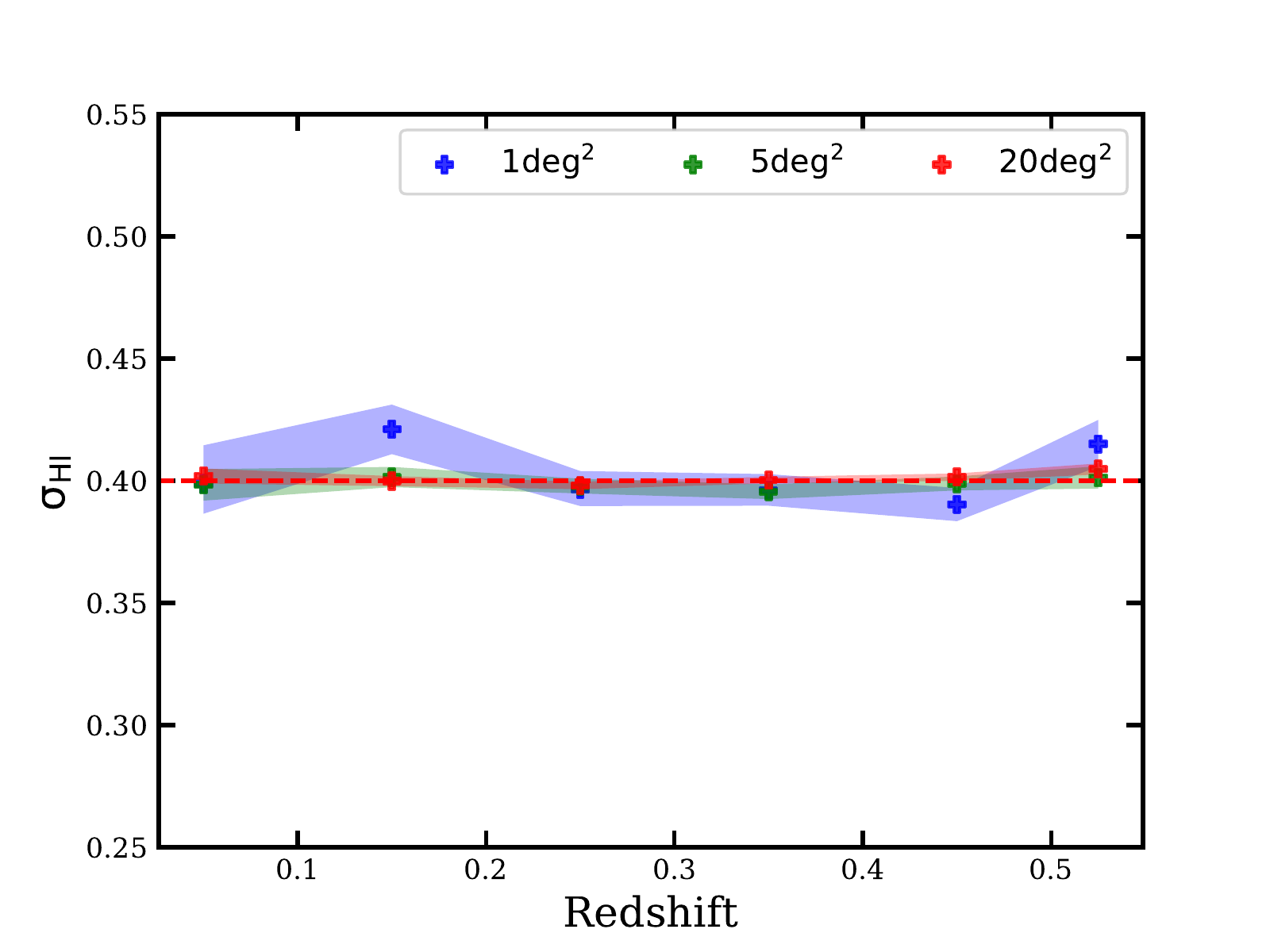}
  \end{subfigure}%
  \hfill
  \begin{subfigure}[b]{0.33\textwidth}
    \includegraphics[width=1.1\columnwidth, height=0.86\columnwidth]{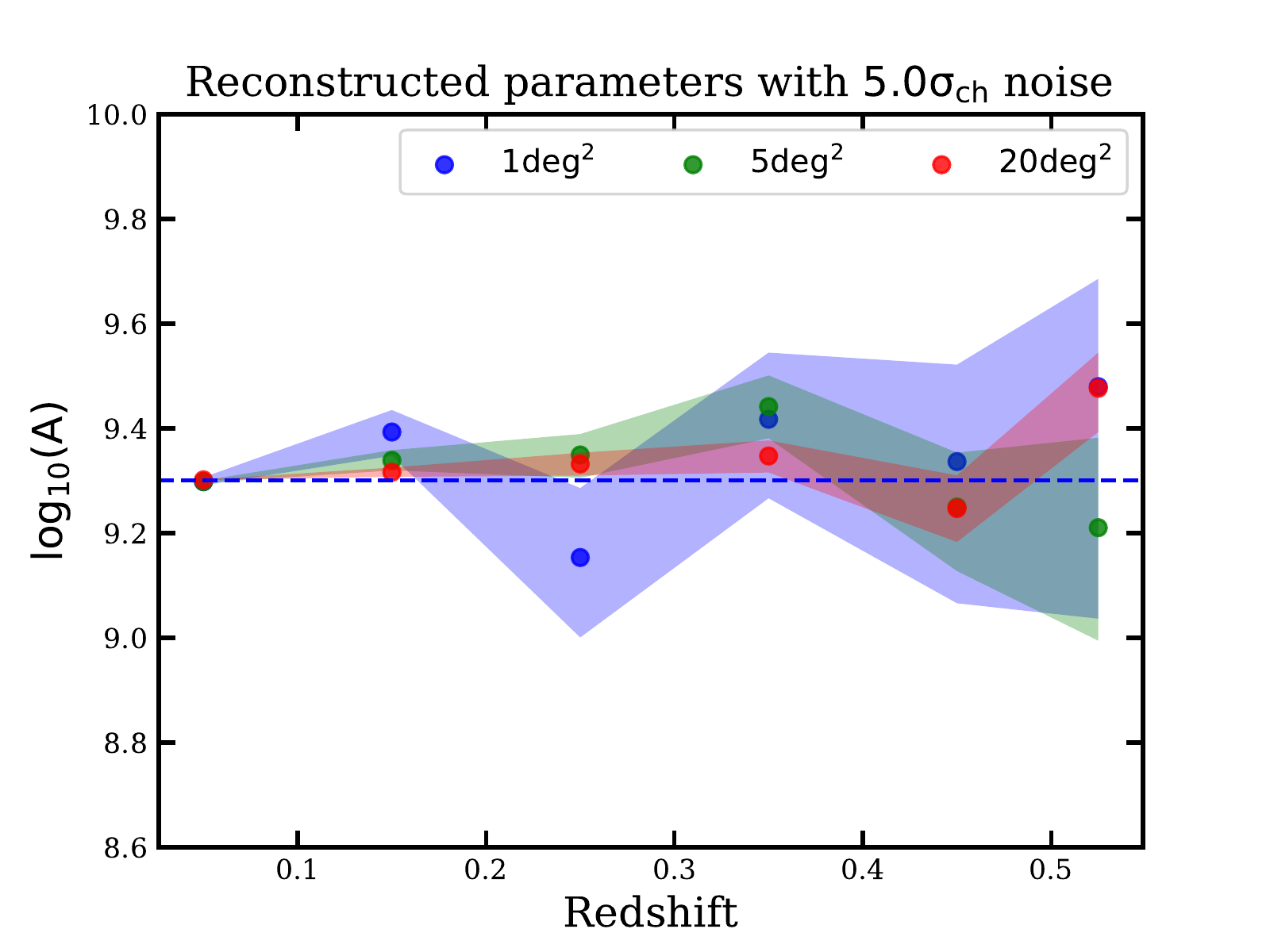}
    \includegraphics[width=1.1\columnwidth, height=0.86\columnwidth]{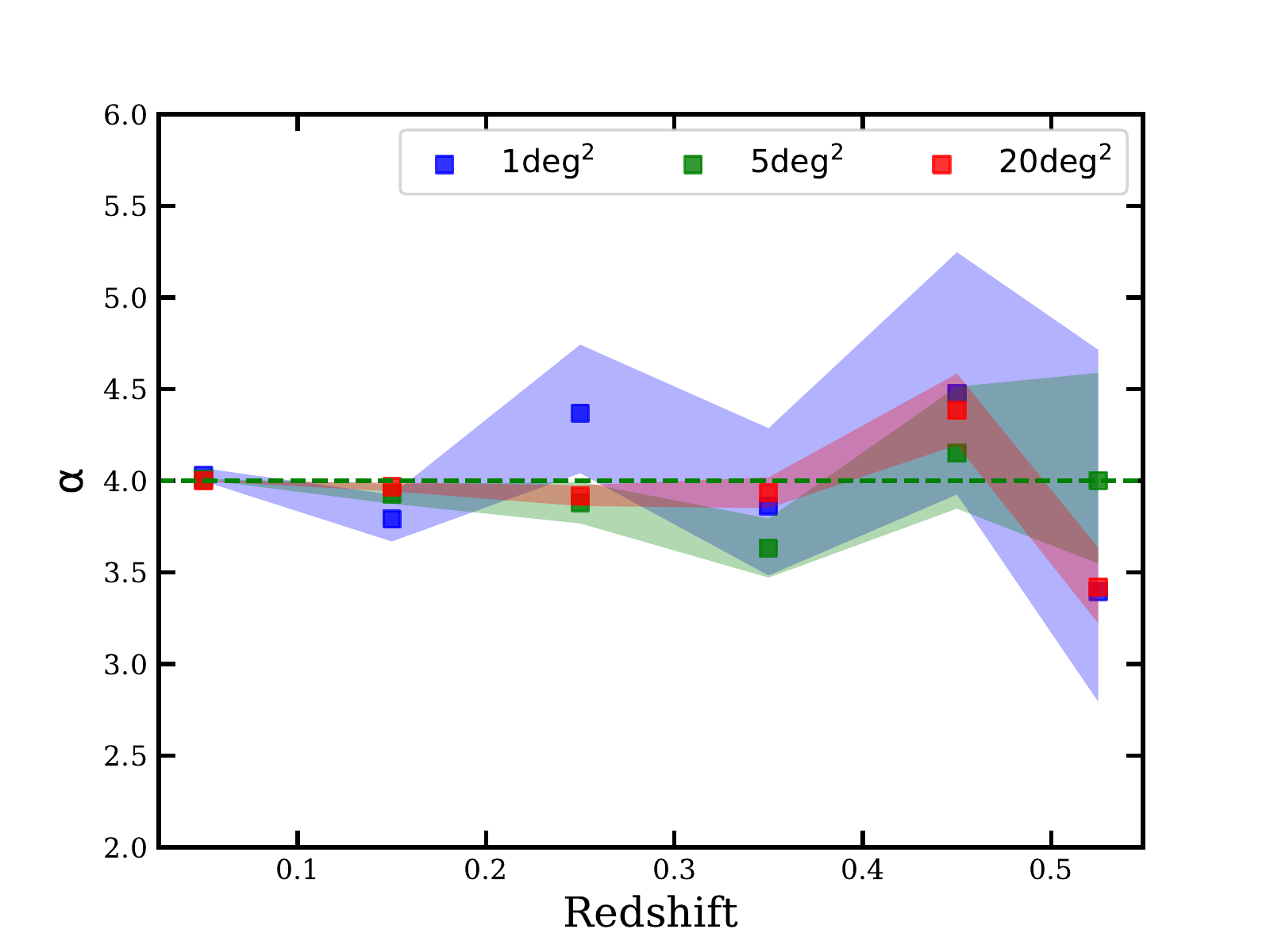}
    \includegraphics[width=1.1\columnwidth, height=0.86\columnwidth]{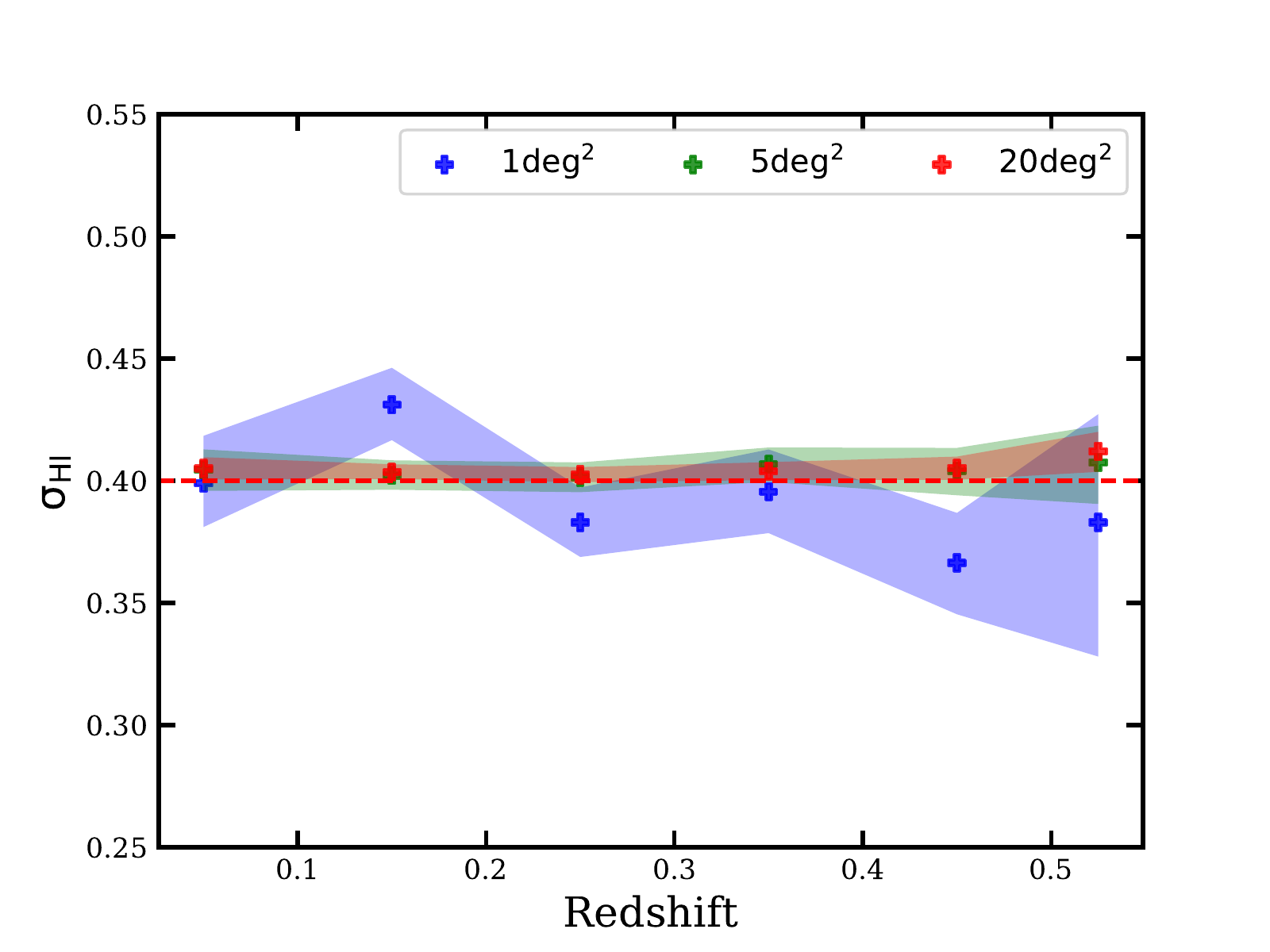}
  \end{subfigure}%
 \caption{Reconstructed parameters of our bTFr model from the simulated surveys as a function of redshift. Left: reconstructed from 0.5$\sigma_{\rm ch}$ noise. Middle: 1$\sigma_{\rm ch}$ noise. Right: 5$\sigma_{\rm ch}$ noise.}\label{fig:pars_160}
\end{figure*}

Our ability to constrain the redshift dependence of $\sigma_{\rm HI}$ is consistent with P20 (i.e. the strongest constraints on the $M_{\rm HI}-M_{\star}$ relation are provided by the middle redshift range). This is due to the fact that the power of Bayesian stacking technique relies on data in the sweet-spot of having a sufficiently large number of galaxies that dominate over the noise at a given integrated flux limit. However, the constraints on the bTFr parameters (i.e. $A$ and $\alpha$) are getting stronger at lower redshifts as the bTFr parameters are very relevant to the \ha line profile that are more prominent from nearby sources. We list all the reconstructed parameters across the redshift bins in Table~\ref{tab:pars_160}, and also represent them in Figure~\ref{fig:pars_160} for a easy visual inspection. If we assume that the bTFr does not evolve with redshift, then we can reconstruct bTFr from the much broader redshift range defined by the limit of our observational bandwidth (Figure~\ref{fig:mdisk-w}). This is obviously  better constrained in comparison to the single redshift bins as expected, and the posterior distributions are shown in Figure~\ref{fig:corner}.

We note again that there are five free parameters to be constrained for the simulated \ha samples at $z>0.1$, as we need the extra two parameters for describing the shape of the double horns, and there is no need to marginalise over them due to the large source density in the high-redshift regime, but we can treat them as nuisance variables since we are only interested in measuring the bTFr here, therefore we do not show them in Figure~\ref{fig:corner_160}. Nevertheless, as we demonstrate in Figure~\ref{fig:corner}, the reconstructed nuisance parameters $b$ and $r_{\rm c}$ are also very close to their input ranges, which indicates that our model can also provide constraints of the line steepness and the depth of the trough.

To evaluate the effect of the double-horn profiles on our results, we have also simulated a group of \ha lines with only Gaussian profiles, where the nuisance parameters are not needed. For the same survey setup we find that the uncertainties for the reconstructed parameters $A$, $\alpha$ and $\sigma_{\rm HI}$ increase slightly. This means that the double-horn profiles are in fact contributing positively to measuring the bTFr, since their profiles with steep line flanks are more easily characterised in the noisy flux densities than the Gaussian profiles for our simulations.

\subsection{The effects of changing the noise and survey area}

We also show the reconstructed bTFr from the cubes with $0.5\sigma_{\rm ch}$ and $5\sigma_{\rm ch}$ Gaussian noise in the left and right panels of Figure~\ref{fig:mdisk-w_160}, and with a range of survey areas from top to bottom panels. This layout is the same for Figure~\ref{fig:corner_160}.

The obvious effect of the increased noise is shown with the blue area of the top right panel of Figure~\ref{fig:mdisk-w_160}, where the statistical uncertainties increase significantly, following the increase in noise by a factor of five. This also has an impact on the redshift range that best constrains $\sigma_{\rm HI}$ for the $M_{\rm HI}-M_{\star}$ relation, but the effect on the redshift dependence of the bTFr parameters are less significant, since the ratio of signal to noise is greatly reduced at all redshifts.

Increasing the survey area has great benefits for measuring the bTFr accurately, especially in the presence of a strong noise as the increase in number of galaxies compensates for the increase in instrumental noise, i.e. the effect of $\sqrt{N}$. At the higher redshifts, the $5\sigma_{\rm ch}$ noise significantly reduces the overall signal-to-noise ratio, to the point where even the largest survey area cannot remedy this impact. Nevertheless, the increase of the survey area by a factor of 5 clearly provides better (about twice stronger on average) constraints for all the parameters as indicated most clearly by the last column of Table~\ref{tab:pars_160} where we take into account all the single z-bins at $z < 0.3$ together.

\section{Conclusions}
\label{sec:conclusions}

We present a 2D flux density model for the observed \ha emission lines from galaxies and a Bayesian stacking technique for measuring the baryonic Tully-Fisher relation and the $M_{\rm HI} - M_{\star}$ relation over the redshift range 0 < z < 0.55, below the detection threshold and down to $M_{\star}=10^{7}$\,M$_\odot$. We simulate the galaxy catalogues and \ha cubes with a range of noise properties and survey areas, and we find that:

\begin{itemize}

\item Our model can reproduce the input bTFr parameters most accurately in a redshfit range from the nearby Universe to $z = 0.3$ for all levels of noise and sky areas, and up to $z = 0.55$ for the nominal level of noise and the wider areas than that of the baseline survey. This is due to the strong signal-to-noise ratio (SNR) of \ha lines, and the number of galaxies covering a wide mass range with a MIGHTEE-like survey, which is important to constrain the normalisation and slope of the bTFr.

\item The strongest constraints on the $M_{\rm HI}-M_{\star}$relation are provided by the middle redshift range ($0.2 < z < 0.3$) due to a sufficiently large number of galaxies that dominate over the noise at a given integrated flux limit.

\item Our model for recovering the bTFr does not perform as well with higher noise at higher redshifts ($z > 0.3$) as expected. This is predominantly due to the reduced number of sources that lie just above the nominal noise threshold.

\item With a variety of shapes of the double-horn profiles, our technique shows an excellent potential for measuring the bTFr and the $M_{\rm HI}-M_{\star}$ relation at high redshifts, where the direct measurements are not possible due to the intrinsic faintness of the \ha signal. Moreover, the recovery of the velocity and \ha mass of the low mass galaxies will be a major step forward in better constraining the statistical properties of the bTFr. Our model can also predict the distribution in the shape of \ha emission lines below the detection threshold, and shed light on the dynamics of galaxies, which would remain undetected otherwise. When we apply our technique to real data it will be possible to perform direct comparisons between the observations and cosmological simulations of galaxy formation and evolution beyond the local Universe, and for the first time in the low mass regime.

\end{itemize}

Finally, it is important to note that since our approach for measuring the bTFr is in fact based on the Bayesian stacking technique used in P20, this approach will inevitably inherit the similar limitations listed at the end of P20 such as dealing with the real noise and instrumental effects in radio observations. We will address these by applying our novel technique to the real MIGHTEE survey dataset in our upcoming work. 

\section*{acknowledgements}
We are grateful to the anonymous referee for helpful comments that have improved this paper. HP, MJJ and MGS acknowledge support from the South African Radio Astronomy Observatory (SARAO) towards this research (www.sarao.ac.za). MGS also acknowledges support from the National Research Foundation (Grant No.~84156). This work was also supported by the CSC of China and the STFC of the United Kingdom. AAP acknowledges the support of the STFC consolidated grant
ST/S000488/1. NM acknowledges support from the Bundesministerium f{\"u}r Bildung und Forschung (BMBF) award 05A20WM4.

\section*{Data availability}
All data used in this paper is available on request to the corresponding authors.

\bibliographystyle{mnras}
\bibliography{references}

\renewcommand{\arraystretch}{1.4}

\begin{table*}
	\centering
	\caption{Reconstructed parameters of our bTFr model from 1, 5 and 20 $\rm deg^2$ surveys with changing noise. Note that we only show the results from combining the \ha samples at $ 0 < z < 0.3$ as we find that this redshift range provides the strongest constraints on our model.}
	\label{tab:pars_160}
	\begin{tabular}{lrrrrrrrr} 
	\hline
	\hline
	Parameter & $\times \; \sigma_{\rm ch} $ & $z < 0.1$ & $ 0.1 < z < 0.2$ & $ 0.2<z < 0.3$ & $ 0.3<z < 0.4$ & $ 0.4<z < 0.5$ & $ 0.5<z < 0.55$ & $ 0 < z < 0.3$ \\
	\hline
	1 $\rm deg^2$ \\
	\hline

        \input{Tables/meerkat_simu_pars_01_160_0.4_00.txt}
	\hline
        \input{Tables/meerkat_simu_pars_01_160_0.4_01.txt}
	\hline
        \input{Tables/meerkat_simu_pars_01_160_0.4_02.txt}
	\hline
	5 $\rm deg^2$ \\
	\hline
        \input{Tables/meerkat_simu_pars_05_160_0.4_00.txt}
	\hline
        \input{Tables/meerkat_simu_pars_05_160_0.4_01.txt}
	\hline
        \input{Tables/meerkat_simu_pars_05_160_0.4_02.txt}
	\hline
	20 $\rm deg^2$ \\
	\hline
        \input{Tables/meerkat_simu_pars_20_160_0.4_00.txt}
	\hline
        \input{Tables/meerkat_simu_pars_20_160_0.4_01.txt}
	\hline
        \input{Tables/meerkat_simu_pars_20_160_0.4_02.txt}
	\hline
	\end{tabular}
\end{table*}







\bsp	
\label{lastpage}
\end{document}

%% file: Tables/meerkat_simu_pars_01_160_0.4_00.txt
$\log_{10}(A)$ & 0.5 & 9.301$^{+0.001}_{-0.001}$ & 9.302$^{+0.003}_{-0.004}$ & 9.305$^{+0.008}_{-0.007}$ & 9.293$^{+0.013}_{-0.008}$ & 9.323$^{+0.015}_{-0.016}$ & 9.363$^{+0.033}_{-0.031}$ & 9.299$^{+0.001}_{-0.000}$ \\
$\log_{10}(A)$ & 1.0 & 9.294$^{+0.001}_{-0.001}$ & 9.303$^{+0.007}_{-0.007}$ & 9.292$^{+0.015}_{-0.016}$ & 9.271$^{+0.027}_{-0.033}$ & 9.338$^{+0.041}_{-0.042}$ & 9.447$^{+0.055}_{-0.052}$ & 9.294$^{+0.001}_{-0.001}$ \\
$\log_{10}(A)$ & 5.0 & 9.298$^{+0.008}_{-0.007}$ & 9.393$^{+0.042}_{-0.045}$ & 9.153$^{+0.133}_{-0.153}$ & 9.417$^{+0.127}_{-0.151}$ & 9.337$^{+0.185}_{-0.271}$ & 9.48$^{+0.206}_{-0.443}$ & 9.3$^{+0.006}_{-0.008}$ \\

%% file: Tables/meerkat_simu_pars_01_160_0.4_01.txt
$\alpha$ & 0.5 & 4.003$^{+0.003}_{-0.004}$ & 4.009$^{+0.012}_{-0.009}$ & 3.983$^{+0.02}_{-0.02}$ & 4.028$^{+0.03}_{-0.038}$ & 3.925$^{+0.046}_{-0.04}$ & 3.884$^{+0.087}_{-0.082}$ & 4.024$^{+0.001}_{-0.001}$ \\
$\alpha$ & 1.0 & 4.021$^{+0.006}_{-0.005}$ & 4.01$^{+0.02}_{-0.021}$ & 4.025$^{+0.04}_{-0.039}$ & 4.113$^{+0.089}_{-0.076}$ & 3.938$^{+0.103}_{-0.105}$ & 3.702$^{+0.148}_{-0.139}$ & 4.033$^{+0.002}_{-0.003}$ \\
$\alpha$ & 5.0 & 4.03$^{+0.038}_{-0.032}$ & 3.791$^{+0.133}_{-0.122}$ & 4.369$^{+0.374}_{-0.328}$ & 3.863$^{+0.424}_{-0.379}$ & 4.476$^{+0.77}_{-0.551}$ & 3.395$^{+1.32}_{-0.6}$ & 4.023$^{+0.013}_{-0.015}$ \\

%% file: Tables/meerkat_simu_pars_01_160_0.4_02.txt
$\sigma_{\rm HI}$ & 0.5 & 0.384$^{+0.012}_{-0.013}$ & 0.419$^{+0.009}_{-0.009}$ & 0.398$^{+0.006}_{-0.006}$ & 0.392$^{+0.005}_{-0.005}$ & 0.397$^{+0.005}_{-0.005}$ & 0.407$^{+0.007}_{-0.007}$ & 0.404$^{+0.004}_{-0.003}$ \\
$\sigma_{\rm HI}$ & 1.0 & 0.399$^{+0.016}_{-0.012}$ & 0.421$^{+0.01}_{-0.01}$ & 0.397$^{+0.007}_{-0.007}$ & 0.396$^{+0.007}_{-0.006}$ & 0.39$^{+0.007}_{-0.007}$ & 0.415$^{+0.01}_{-0.009}$ & 0.405$^{+0.006}_{-0.004}$ \\
$\sigma_{\rm HI}$ & 5.0 & 0.399$^{+0.019}_{-0.018}$ & 0.431$^{+0.015}_{-0.014}$ & 0.383$^{+0.014}_{-0.014}$ & 0.396$^{+0.017}_{-0.017}$ & 0.366$^{+0.02}_{-0.021}$ & 0.383$^{+0.044}_{-0.055}$ & 0.408$^{+0.009}_{-0.007}$ \\

%% file: Tables/meerkat_simu_pars_05_160_0.4_00.txt
$\log_{10}(A)$ & 0.5 & 9.299$^{+0.000}_{-0.000}$ & 9.305$^{+0.001}_{-0.001}$ & 9.31$^{+0.003}_{-0.003}$ & 9.293$^{+0.004}_{-0.005}$ & 9.321$^{+0.009}_{-0.007}$ & 9.327$^{+0.016}_{-0.016}$ & 9.3$^{+0.000}_{-0.000}$ \\
$\log_{10}(A)$ & 1.0 & 9.299$^{+0.001}_{-0.000}$ & 9.306$^{+0.003}_{-0.003}$ & 9.301$^{+0.006}_{-0.007}$ & 9.292$^{+0.01}_{-0.012}$ & 9.321$^{+0.015}_{-0.024}$ & 9.35$^{+0.031}_{-0.036}$ & 9.299$^{+0.000}_{-0.000}$ \\
$\log_{10}(A)$ & 5.0 & 9.298$^{+0.002}_{-0.002}$ & 9.339$^{+0.02}_{-0.019}$ & 9.349$^{+0.04}_{-0.044}$ & 9.442$^{+0.059}_{-0.061}$ & 9.25$^{+0.104}_{-0.122}$ & 9.21$^{+0.172}_{-0.216}$ & 9.3$^{+0.002}_{-0.003}$ \\

%% file: Tables/meerkat_simu_pars_05_160_0.4_01.txt
$\alpha$ & 0.5 & 4.007$^{+0.000}_{-0.000}$ & 3.999$^{+0.004}_{-0.003}$ & 3.985$^{+0.009}_{-0.009}$ & 4.032$^{+0.014}_{-0.014}$ & 3.952$^{+0.021}_{-0.022}$ & 3.947$^{+0.044}_{-0.044}$ & 4.007$^{+0.000}_{-0.000}$ \\
$\alpha$ & 1.0 & 4.008$^{+0.001}_{-0.001}$ & 3.993$^{+0.009}_{-0.009}$ & 4.009$^{+0.02}_{-0.019}$ & 4.06$^{+0.032}_{-0.031}$ & 3.955$^{+0.051}_{-0.044}$ & 3.886$^{+0.086}_{-0.082}$ & 4.008$^{+0.000}_{-0.001}$ \\
$\alpha$ & 5.0 & 4.009$^{+0.003}_{-0.003}$ & 3.928$^{+0.054}_{-0.054}$ & 3.879$^{+0.114}_{-0.112}$ & 3.632$^{+0.162}_{-0.16}$ & 4.151$^{+0.362}_{-0.303}$ & 4.0$^{+0.588}_{-0.452}$ & 4.007$^{+0.004}_{-0.002}$ \\

%% file: Tables/meerkat_simu_pars_05_160_0.4_02.txt
$\sigma_{\rm HI}$ & 0.5 & 0.398$^{+0.007}_{-0.005}$ & 0.398$^{+0.004}_{-0.004}$ & 0.398$^{+0.003}_{-0.003}$ & 0.397$^{+0.002}_{-0.002}$ & 0.399$^{+0.002}_{-0.002}$ & 0.401$^{+0.003}_{-0.003}$ & 0.4$^{+0.001}_{-0.002}$ \\
$\sigma_{\rm HI}$ & 1.0 & 0.398$^{+0.006}_{-0.007}$ & 0.401$^{+0.004}_{-0.004}$ & 0.398$^{+0.003}_{-0.003}$ & 0.395$^{+0.003}_{-0.003}$ & 0.399$^{+0.003}_{-0.003}$ & 0.401$^{+0.005}_{-0.005}$ & 0.402$^{+0.002}_{-0.002}$ \\
$\sigma_{\rm HI}$ & 5.0 & 0.405$^{+0.008}_{-0.009}$ & 0.402$^{+0.006}_{-0.006}$ & 0.401$^{+0.006}_{-0.006}$ & 0.407$^{+0.007}_{-0.007}$ & 0.404$^{+0.009}_{-0.01}$ & 0.407$^{+0.015}_{-0.017}$ & 0.404$^{+0.004}_{-0.003}$ \\

%% file: Tables/meerkat_simu_pars_20_160_0.4_00.txt
$\log_{10}(A)$ & 0.5 & 9.302$^{+0.000}_{-0.000}$ & 9.307$^{+0.001}_{-0.001}$ & 9.303$^{+0.004}_{-0.001}$ & 9.298$^{+0.005}_{-0.003}$ & 9.316$^{+0.005}_{-0.005}$ & 9.308$^{+0.01}_{-0.007}$ & 9.302$^{+0.000}_{-0.000}$ \\
$\log_{10}(A)$ & 1.0 & 9.302$^{+0.000}_{-0.000}$ & 9.303$^{+0.001}_{-0.001}$ & 9.302$^{+0.003}_{-0.003}$ & 9.319$^{+0.004}_{-0.007}$ & 9.321$^{+0.008}_{-0.011}$ & 9.32$^{+0.014}_{-0.014}$ & 9.302$^{+0.000}_{-0.000}$ \\
$\log_{10}(A)$ & 5.0 & 9.302$^{+0.000}_{-0.000}$ & 9.316$^{+0.009}_{-0.009}$ & 9.332$^{+0.021}_{-0.022}$ & 9.347$^{+0.03}_{-0.032}$ & 9.247$^{+0.063}_{-0.064}$ & 9.477$^{+0.068}_{-0.084}$ & 9.302$^{+0.000}_{-0.000}$ \\

%% file: Tables/meerkat_simu_pars_20_160_0.4_01.txt
$\alpha$ & 0.5 & 4.0$^{+0.000}_{-0.000}$ & 3.992$^{+0.002}_{-0.002}$ & 4.0$^{+0.004}_{-0.008}$ & 4.022$^{+0.009}_{-0.012}$ & 3.97$^{+0.012}_{-0.012}$ & 3.983$^{+0.022}_{-0.022}$ & 4.0$^{+0.000}_{-0.000}$ \\
$\alpha$ & 1.0 & 4.0$^{+0.000}_{-0.000}$ & 4.002$^{+0.004}_{-0.004}$ & 4.008$^{+0.008}_{-0.009}$ & 3.96$^{+0.017}_{-0.013}$ & 3.957$^{+0.026}_{-0.024}$ & 3.945$^{+0.041}_{-0.042}$ & 4.0$^{+0.000}_{-0.000}$ \\
$\alpha$ & 5.0 & 3.999$^{+0.001}_{-0.001}$ & 3.967$^{+0.023}_{-0.025}$ & 3.917$^{+0.058}_{-0.056}$ & 3.934$^{+0.084}_{-0.083}$ & 4.384$^{+0.2}_{-0.19}$ & 3.42$^{+0.214}_{-0.199}$ & 3.999$^{+0.001}_{-0.001}$ \\

%% file: Tables/meerkat_simu_pars_20_160_0.4_02.txt
$\sigma_{\rm HI}$ & 0.5 & 0.398$^{+0.004}_{-0.002}$ & 0.4$^{+0.002}_{-0.002}$ & 0.398$^{+0.001}_{-0.001}$ & 0.399$^{+0.001}_{-0.001}$ & 0.401$^{+0.001}_{-0.001}$ & 0.403$^{+0.002}_{-0.001}$ & 0.402$^{+0.001}_{-0.001}$ \\
$\sigma_{\rm HI}$ & 1.0 & 0.402$^{+0.003}_{-0.003}$ & 0.4$^{+0.002}_{-0.002}$ & 0.398$^{+0.001}_{-0.002}$ & 0.4$^{+0.001}_{-0.001}$ & 0.402$^{+0.001}_{-0.002}$ & 0.405$^{+0.002}_{-0.002}$ & 0.401$^{+0.001}_{-0.001}$ \\
$\sigma_{\rm HI}$ & 5.0 & 0.405$^{+0.005}_{-0.005}$ & 0.404$^{+0.003}_{-0.003}$ & 0.402$^{+0.003}_{-0.003}$ & 0.404$^{+0.004}_{-0.004}$ & 0.405$^{+0.005}_{-0.005}$ & 0.412$^{+0.008}_{-0.008}$ & 0.404$^{+0.002}_{-0.002}$ \\

%% file: mnras_tully_fisher_hengxing_mjj.bbl
\begin{thebibliography}{}
\makeatletter
\relax
\def\mn@urlcharsother{\let\do\@makeother \do\$\do\&\do\#\do\^\do\_\do\%\do\~}
\def\mn@doi{\begingroup\mn@urlcharsother \@ifnextchar [ {\mn@doi@}
  {\mn@doi@[]}}
\def\mn@doi@[#1]#2{\def\@tempa{#1}\ifx\@tempa\@empty \href
  {http://dx.doi.org/#2} {doi:#2}\else \href {http://dx.doi.org/#2} {#1}\fi
  \endgroup}
\def\mn@eprint#1#2{\mn@eprint@#1:#2::\@nil}
\def\mn@eprint@arXiv#1{\href {http://arxiv.org/abs/#1} {{\tt arXiv:#1}}}
\def\mn@eprint@dblp#1{\href {http://dblp.uni-trier.de/rec/bibtex/#1.xml}
  {dblp:#1}}
\def\mn@eprint@#1:#2:#3:#4\@nil{\def\@tempa {#1}\def\@tempb {#2}\def\@tempc
  {#3}\ifx \@tempc \@empty \let \@tempc \@tempb \let \@tempb \@tempa \fi \ifx
  \@tempb \@empty \def\@tempb {arXiv}\fi \@ifundefined
  {mn@eprint@\@tempb}{\@tempb:\@tempc}{\expandafter \expandafter \csname
  mn@eprint@\@tempb\endcsname \expandafter{\@tempc}}}

\bibitem[\protect\citeauthoryear{Abril-Melgarejo et~al.,}{Abril-Melgarejo
  et~al.}{2021}]{Abril-Melgarejo2021}
Abril-Melgarejo V.,  et~al., 2021, The Tully-Fisher relation in dense groups at
  $z \sim 0.7$ in the MAGIC survey (\mn@eprint {arXiv} {2101.08069})

\bibitem[\protect\citeauthoryear{{Arnett}}{{Arnett}}{1996}]{A1996book}
{Arnett} D.,  1996, {Supernovae and Nucleosynthesis: An Investigation of the
  History of Matter from the Big Bang to the Present}

\bibitem[\protect\citeauthoryear{{Bedregal}, {Arag{\'o}n-Salamanca}  \&
  {Merrifield}}{{Bedregal} et~al.}{2006}]{Bedregal2006}
{Bedregal} A.~G.,  {Arag{\'o}n-Salamanca} A.,   {Merrifield} M.~R.,  2006,
  \mn@doi [\mnras] {10.1111/j.1365-2966.2006.11031.x}, \href
  {https://ui.adsabs.harvard.edu/abs/2006MNRAS.373.1125B} {373, 1125}

\bibitem[\protect\citeauthoryear{{Begum}, {Chengalur}, {Karachentsev}  \&
  {Sharina}}{{Begum} et~al.}{2008}]{Begum2008}
{Begum} A.,  {Chengalur} J.~N.,  {Karachentsev} I.~D.,   {Sharina} M.~E.,
  2008, \mn@doi [\mnras] {10.1111/j.1365-2966.2008.13010.x}, \href
  {https://ui.adsabs.harvard.edu/abs/2008MNRAS.386..138B} {386, 138}

\bibitem[\protect\citeauthoryear{{Bradford}, {Geha}  \& {van den
  Bosch}}{{Bradford} et~al.}{2016}]{bradford16}
{Bradford} J.~D.,  {Geha} M.~C.,   {van den Bosch} F.~C.,  2016, \mn@doi [\apj]
  {10.3847/0004-637X/832/1/11}, \href
  {https://ui.adsabs.harvard.edu/abs/2016ApJ...832...11B} {832, 11}

\bibitem[\protect\citeauthoryear{{Buchner} et~al.,}{{Buchner}
  et~al.}{2014}]{buchner2014x}
{Buchner} J.,  et~al., 2014, \mn@doi [\aap] {10.1051/0004-6361/201322971},
  \href {https://ui.adsabs.harvard.edu/abs/2014A&A...564A.125B} {564, A125}

\bibitem[\protect\citeauthoryear{Chowdhury, Kanekar, Chengalur, Sethi  \&
  Dwarakanath}{Chowdhury et~al.}{2020}]{Chowdhury_2020}
Chowdhury A.,  Kanekar N.,  Chengalur J.~N.,  Sethi S.,   Dwarakanath K.~S.,
  2020, \mn@doi [Nature] {10.1038/s41586-020-2794-7}, 586, 369–372

\bibitem[\protect\citeauthoryear{{Chung}, {van Gorkom}, {O'Neil}  \&
  {Bothun}}{{Chung} et~al.}{2002}]{Chung2002}
{Chung} A.,  {van Gorkom} J.~H.,  {O'Neil} K.,   {Bothun} G.~D.,  2002, \mn@doi
  [\aj] {10.1086/339979}, \href
  {https://ui.adsabs.harvard.edu/abs/2002AJ....123.2387C} {123, 2387}

\bibitem[\protect\citeauthoryear{{Courteau}, {Andersen}, {Bershady},
  {MacArthur}  \& {Rix}}{{Courteau} et~al.}{2003}]{courteau03}
{Courteau} S.,  {Andersen} D.~R.,  {Bershady} M.~A.,  {MacArthur} L.~A.,
  {Rix} H.-W.,  2003, \mn@doi [\apj] {10.1086/376754}, \href
  {https://ui.adsabs.harvard.edu/abs/2003ApJ...594..208C} {594, 208}

\bibitem[\protect\citeauthoryear{{Dav{\'e}}, {Angl{\'e}s-Alc{\'a}zar},
  {Narayanan}, {Li}, {Rafieferantsoa}  \& {Appleby}}{{Dav{\'e}}
  et~al.}{2019}]{Dave2019}
{Dav{\'e}} R.,  {Angl{\'e}s-Alc{\'a}zar} D.,  {Narayanan} D.,  {Li} Q.,
  {Rafieferantsoa} M.~H.,   {Appleby} S.,  2019, \mn@doi [\mnras]
  {10.1093/mnras/stz937}, \href
  {https://ui.adsabs.harvard.edu/abs/2019MNRAS.486.2827D} {486, 2827}

\bibitem[\protect\citeauthoryear{{Delhaize}, {Meyer}, {Staveley-Smith}  \&
  {Boyle}}{{Delhaize} et~al.}{2013}]{delhaize2013detection}
{Delhaize} J.,  {Meyer} M.~J.,  {Staveley-Smith} L.,   {Boyle} B.~J.,  2013,
  \mn@doi [\mnras] {10.1093/mnras/stt810}, \href
  {https://ui.adsabs.harvard.edu/abs/2013MNRAS.433.1398D} {433, 1398}

\bibitem[\protect\citeauthoryear{{Desmond}}{{Desmond}}{2012}]{Desmond2012}
{Desmond} H.,  2012, arXiv e-prints, \href
  {https://ui.adsabs.harvard.edu/abs/2012arXiv1204.1497D} {p. arXiv:1204.1497}

\bibitem[\protect\citeauthoryear{Dubois et~al.,}{Dubois
  et~al.}{2020}]{Dubois2020}
Dubois Y.,  et~al., 2020, Introducing the NewHorizon simulation: galaxy
  properties with resolved internal dynamics across cosmic time (\mn@eprint
  {arXiv} {2009.10578})

\bibitem[\protect\citeauthoryear{{Fern{\'a}ndez} et~al.,}{{Fern{\'a}ndez}
  et~al.}{2016}]{fernandez2016}
{Fern{\'a}ndez} X.,  et~al., 2016, \mn@doi [ApJ] {10.3847/2041-8205/824/1/L1},
  \href {https://ui.adsabs.harvard.edu/abs/2016ApJ...824L...1F} {824, L1}

\bibitem[\protect\citeauthoryear{{Feroz}, {Hobson}  \& {Bridges}}{{Feroz}
  et~al.}{2009}]{feroz2009multinest}
{Feroz} F.,  {Hobson} M.~P.,   {Bridges} M.,  2009, \mn@doi [\mnras]
  {10.1111/j.1365-2966.2009.14548.x}, \href
  {https://ui.adsabs.harvard.edu/abs/2009MNRAS.398.1601F} {398, 1601}

\bibitem[\protect\citeauthoryear{Glowacki, Elson  \& Davé}{Glowacki
  et~al.}{2020a}]{Glowacki2020}
Glowacki M.,  Elson E.,   Davé R.,  2020a, The redshift evolution of the
  baryonic Tully-Fisher relation in Simba (\mn@eprint {arXiv} {2011.08866})

\bibitem[\protect\citeauthoryear{{Glowacki}, {Elson}  \& {Dav{\'e}}}{{Glowacki}
  et~al.}{2020b}]{glowacki20}
{Glowacki} M.,  {Elson} E.,   {Dav{\'e}} R.,  2020b, \mn@doi [\mnras]
  {10.1093/mnras/staa2616}, \href
  {https://ui.adsabs.harvard.edu/abs/2020MNRAS.498.3687G} {498, 3687}

\bibitem[\protect\citeauthoryear{{Iorio}, {Fraternali}, {Nipoti}, {Di Teodoro},
  {Read}  \& {Battaglia}}{{Iorio} et~al.}{2017}]{iorio2017}
{Iorio} G.,  {Fraternali} F.,  {Nipoti} C.,  {Di Teodoro} E.,  {Read} J.~I.,
  {Battaglia} G.,  2017, \mn@doi [\mnras] {10.1093/mnras/stw3285}, \href
  {https://ui.adsabs.harvard.edu/abs/2017MNRAS.466.4159I} {466, 4159}

\bibitem[\protect\citeauthoryear{{Jarvis} et~al.,}{{Jarvis}
  et~al.}{2016}]{jarvis2017meerkat}
{Jarvis} M.,  et~al., 2016, in Proceedings of MeerKAT Science: On the Pathway
  to the SKA. 25-27 May. p.~6 (\mn@eprint {arXiv} {1709.01901})

\bibitem[\protect\citeauthoryear{{Jonas} \& {MeerKAT Team}}{{Jonas} \& {MeerKAT
  Team}}{2016}]{Jonas2016}
{Jonas} J.,  {MeerKAT Team} 2016, in Proceedings of MeerKAT Science: On the
  Pathway to the SKA. 25-27 May. p.~1

\bibitem[\protect\citeauthoryear{{Karachentsev}, {Kaisina}  \& {Kashibadze
  Nasonova}}{{Karachentsev} et~al.}{2017}]{Karachentsev2016}
{Karachentsev} I.~D.,  {Kaisina} E.~I.,   {Kashibadze Nasonova} O.~G.,  2017,
  \mn@doi [\aj] {10.3847/1538-3881/153/1/6}, \href
  {https://ui.adsabs.harvard.edu/abs/2017AJ....153....6K} {153, 6}

\bibitem[\protect\citeauthoryear{{Lelli}, {McGaugh}, {Schombert}, {Desmond}  \&
  {Katz}}{{Lelli} et~al.}{2019}]{Lelli19}
{Lelli} F.,  {McGaugh} S.~S.,  {Schombert} J.~M.,  {Desmond} H.,   {Katz} H.,
  2019, \mn@doi [\mnras] {10.1093/mnras/stz205}, \href
  {https://ui.adsabs.harvard.edu/abs/2019MNRAS.484.3267L} {484, 3267}

\bibitem[\protect\citeauthoryear{{Maddox}, {Hess}, {Obreschkow}, {Jarvis}  \&
  {Blyth}}{{Maddox} et~al.}{2015}]{Maddox2015}
{Maddox} N.,  {Hess} K.~M.,  {Obreschkow} D.,  {Jarvis} M.~J.,   {Blyth} S.~L.,
   2015, \mn@doi [\mnras] {10.1093/mnras/stu2532}, \href
  {https://ui.adsabs.harvard.edu/abs/2015MNRAS.447.1610M} {447, 1610}

\bibitem[\protect\citeauthoryear{Maddox et~al.,}{Maddox
  et~al.}{2021}]{Maddox2020}
Maddox N.,  et~al., 2021, \mn@doi [Astronomy & Astrophysics]
  {10.1051/0004-6361/202039655}, 646, A35

\bibitem[\protect\citeauthoryear{{McGaugh}, {Schombert}, {Bothun}  \& {de
  Blok}}{{McGaugh} et~al.}{2000}]{McGaugh2000}
{McGaugh} S.~S.,  {Schombert} J.~M.,  {Bothun} G.~D.,   {de Blok} W.~J.~G.,
  2000, \mn@doi [\apjl] {10.1086/312628}, \href
  {https://ui.adsabs.harvard.edu/abs/2000ApJ...533L..99M} {533, L99}

\bibitem[\protect\citeauthoryear{{Meyer}, {Meyer}, {Obreschkow}  \&
  {Staveley-Smith}}{{Meyer} et~al.}{2016}]{Meyer2016}
{Meyer} S.~A.,  {Meyer} M.,  {Obreschkow} D.,   {Staveley-Smith} L.,  2016,
  \mn@doi [\mnras] {10.1093/mnras/stv2458}, \href
  {https://ui.adsabs.harvard.edu/abs/2016MNRAS.455.3136M} {455, 3136}

\bibitem[\protect\citeauthoryear{{Meyer}, {Robotham}, {Obreschkow},
  {Westmeier}, {Duffy}  \& {Staveley-Smith}}{{Meyer}
  et~al.}{2017}]{meyer2017tracing}
{Meyer} M.,  {Robotham} A.,  {Obreschkow} D.,  {Westmeier} T.,  {Duffy} A.~R.,
   {Staveley-Smith} L.,  2017, \mn@doi [\pasa] {10.1017/pasa.2017.31}, \href
  {https://ui.adsabs.harvard.edu/abs/2017PASA...34...52M} {34}

\bibitem[\protect\citeauthoryear{{Pan}, {Jarvis}, {Allison}, {Heywood},
  {Santos}, {Maddox}, {Frank}  \& {Kang}}{{Pan} et~al.}{2020}]{pan2020}
{Pan} H.,  {Jarvis} M.~J.,  {Allison} J.~R.,  {Heywood} I.,  {Santos} M.~G.,
  {Maddox} N.,  {Frank} B.~S.,   {Kang} X.,  2020, \mn@doi [\mnras]
  {10.1093/mnras/stz3030}, \href
  {https://ui.adsabs.harvard.edu/abs/2020MNRAS.491.1227P} {491, 1227}

\bibitem[\protect\citeauthoryear{{Papastergis}, {Cattaneo}, {Huang},
  {Giovanelli}  \& {Haynes}}{{Papastergis} et~al.}{2012}]{Papastergis2012}
{Papastergis} E.,  {Cattaneo} A.,  {Huang} S.,  {Giovanelli} R.,   {Haynes}
  M.~P.,  2012, \mn@doi [\apj] {10.1088/0004-637X/759/2/138}, \href
  {https://ui.adsabs.harvard.edu/abs/2012ApJ...759..138P} {759, 138}

\bibitem[\protect\citeauthoryear{{Papastergis}, {Adams}  \& {van der
  Hulst}}{{Papastergis} et~al.}{2016}]{Pap2016}
{Papastergis} E.,  {Adams} E.~A.~K.,   {van der Hulst} J.~M.,  2016, \mn@doi
  [\aap] {10.1051/0004-6361/201628410}, \href
  {https://ui.adsabs.harvard.edu/abs/2016A&A...593A..39P} {593, A39}

\bibitem[\protect\citeauthoryear{{Parkash}, {Brown}, {Jarrett}  \&
  {Bonne}}{{Parkash} et~al.}{2018}]{Parkash2018}
{Parkash} V.,  {Brown} M. J.~I.,  {Jarrett} T.~H.,   {Bonne} N.~J.,  2018,
  \mn@doi [\apj] {10.3847/1538-4357/aad3b9}, \href
  {https://ui.adsabs.harvard.edu/abs/2018ApJ...864...40P} {864, 40}

\bibitem[\protect\citeauthoryear{{Planck Collaboration} et~al.,}{{Planck
  Collaboration} et~al.}{2016}]{ade2016planck}
{Planck Collaboration} et~al., 2016, \mn@doi [\aap]
  {10.1051/0004-6361/201525830}, \href
  {https://ui.adsabs.harvard.edu/abs/2016A&A...594A..13P} {594, A13}

\bibitem[\protect\citeauthoryear{{Ponomareva}, {Verheijen}, {Peletier}  \&
  {Bosma}}{{Ponomareva} et~al.}{2017}]{ponomareva17}
{Ponomareva} A.~A.,  {Verheijen} M. A.~W.,  {Peletier} R.~F.,   {Bosma} A.,
  2017, \mn@doi [\mnras] {10.1093/mnras/stx1018}, \href
  {https://ui.adsabs.harvard.edu/abs/2017MNRAS.469.2387P} {469, 2387}

\bibitem[\protect\citeauthoryear{{Ponomareva}, {Verheijen}, {Papastergis},
  {Bosma}  \& {Peletier}}{{Ponomareva} et~al.}{2018}]{ponomareva18}
{Ponomareva} A.~A.,  {Verheijen} M. A.~W.,  {Papastergis} E.,  {Bosma} A.,
  {Peletier} R.~F.,  2018, \mn@doi [\mnras] {10.1093/mnras/stx3066}, \href
  {https://ui.adsabs.harvard.edu/abs/2018MNRAS.474.4366P} {474, 4366}

\bibitem[\protect\citeauthoryear{{Rhee}, {Lah}, {Briggs}, {Chengalur},
  {Colless}, {Willner}, {Ashby}  \& {Le F{\`e}vre}}{{Rhee}
  et~al.}{2018}]{rhee2018}
{Rhee} J.,  {Lah} P.,  {Briggs} F.~H.,  {Chengalur} J.~N.,  {Colless} M.,
  {Willner} S.~P.,  {Ashby} M. L.~N.,   {Le F{\`e}vre} O.,  2018, \mn@doi
  [MNRAS] {10.1093/mnras/stx2461}, \href
  {https://ui.adsabs.harvard.edu/abs/2018MNRAS.473.1879R} {473, 1879}

\bibitem[\protect\citeauthoryear{{Schechter}}{{Schechter}}{1976}]{Schechter1976}
{Schechter} P.,  1976, \mn@doi [\apj] {10.1086/154079}, \href
  {https://ui.adsabs.harvard.edu/abs/1976ApJ...203..297S} {203, 297}

\bibitem[\protect\citeauthoryear{{Sorce}, {Tully}, {Courtois}, {Jarrett},
  {Neill}  \& {Shaya}}{{Sorce} et~al.}{2014}]{sorce14}
{Sorce} J.~G.,  {Tully} R.~B.,  {Courtois} H.~M.,  {Jarrett} T.~H.,  {Neill}
  J.~D.,   {Shaya} E.~J.,  2014, \mn@doi [\mnras] {10.1093/mnras/stu1450},
  \href {https://ui.adsabs.harvard.edu/abs/2014MNRAS.444..527S} {444, 527}

\bibitem[\protect\citeauthoryear{{Tiley} et~al.,}{{Tiley}
  et~al.}{2019}]{tiley19}
{Tiley} A.~L.,  et~al., 2019, \mn@doi [\mnras] {10.1093/mnras/sty2794}, \href
  {https://ui.adsabs.harvard.edu/abs/2019MNRAS.482.2166T} {482, 2166}

\bibitem[\protect\citeauthoryear{{Topal}, {Bureau}, {Tiley}, {Davis}  \&
  {Torii}}{{Topal} et~al.}{2018}]{topal18}
{Topal} S.,  {Bureau} M.,  {Tiley} A.~L.,  {Davis} T.~A.,   {Torii} K.,  2018,
  \mn@doi [\mnras] {10.1093/mnras/sty1617}, \href
  {https://ui.adsabs.harvard.edu/abs/2018MNRAS.479.3319T} {479, 3319}

\bibitem[\protect\citeauthoryear{{Trujillo-Gomez}, {Klypin}, {Primack}  \&
  {Romanowsky}}{{Trujillo-Gomez} et~al.}{2011}]{Trujillo-Gomez2010}
{Trujillo-Gomez} S.,  {Klypin} A.,  {Primack} J.,   {Romanowsky} A.~J.,  2011,
  \mn@doi [\apj] {10.1088/0004-637X/742/1/16}, \href
  {https://ui.adsabs.harvard.edu/abs/2011ApJ...742...16T} {742, 16}

\bibitem[\protect\citeauthoryear{{Tully} \& {Fisher}}{{Tully} \&
  {Fisher}}{1977}]{tf77}
{Tully} R.~B.,  {Fisher} J.~R.,  1977, \aap, \href
  {https://ui.adsabs.harvard.edu/abs/1977A&A....54..661T} {500, 105}

\bibitem[\protect\citeauthoryear{{Tully}, {Courtois}, {Hoffman}  \&
  {Pomar{\`e}de}}{{Tully} et~al.}{2014}]{tully14}
{Tully} R.~B.,  {Courtois} H.,  {Hoffman} Y.,   {Pomar{\`e}de} D.,  2014,
  \mn@doi [\nat] {10.1038/nature13674}, \href
  {https://ui.adsabs.harvard.edu/abs/2014Natur.513...71T} {513, 71}

\bibitem[\protect\citeauthoryear{{{\"U}bler} et~al.,}{{{\"U}bler}
  et~al.}{2017}]{ubler17}
{{\"U}bler} H.,  et~al., 2017, \mn@doi [\apj] {10.3847/1538-4357/aa7558}, \href
  {https://ui.adsabs.harvard.edu/abs/2017ApJ...842..121U} {842, 121}

\bibitem[\protect\citeauthoryear{{Westmeier}, {Jurek}, {Obreschkow},
  {Koribalski}  \& {Staveley-Smith}}{{Westmeier} et~al.}{2014}]{Westmeier2014}
{Westmeier} T.,  {Jurek} R.,  {Obreschkow} D.,  {Koribalski} B.~S.,
  {Staveley-Smith} L.,  2014, \mn@doi [\mnras] {10.1093/mnras/stt2266}, \href
  {https://ui.adsabs.harvard.edu/abs/2014MNRAS.438.1176W} {438, 1176}

\bibitem[\protect\citeauthoryear{{Willick}}{{Willick}}{1999}]{willick99}
{Willick} J.~A.,  1999, \mn@doi [\apj] {10.1086/307108}, \href
  {https://ui.adsabs.harvard.edu/abs/1999ApJ...516...47W} {516, 47}

\bibitem[\protect\citeauthoryear{{Zwaan}, {Meyer}, {Staveley-Smith}  \&
  {Webster}}{{Zwaan} et~al.}{2005}]{zwaan2005hipass}
{Zwaan} M.~A.,  {Meyer} M.~J.,  {Staveley-Smith} L.,   {Webster} R.~L.,  2005,
  \mn@doi [\mnras] {10.1111/j.1745-3933.2005.00029.x}, \href
  {https://ui.adsabs.harvard.edu/abs/2005MNRAS.359L..30Z} {359, L30}

\makeatother
\end{thebibliography}
